\documentclass{aa}
\usepackage{graphicx}
\usepackage{txfonts}
\usepackage{lipsum}
\usepackage{subcaption}             
\usepackage{lscape}           
\usepackage{placeins}
                                
\usepackage{xcolor}

\newcommand{\mesa}{\texttt{MESA}}
\newcommand{\kepler}{\texttt{KEPLER}}
\newcommand{\approxai}{\texttt{Approx}\_\texttt{A56}}
\newcommand{\approxaii}{\texttt{Approx}\_\texttt{A64}}
\newcommand{\middmdot}{$\dot{M}_{\mathrm{moderate}~}$}
\newcommand{\highmdot}{$\dot{M}_{\mathrm{high}~}$}
\newcommand{\hydros}{$^{1}\mathrm{H}$}
\newcommand{\helioos}{$^{4}\mathrm{He}$}
\newcommand{\carbos}{$^{12}\mathrm{C}$}
\newcommand{\neos}{$^{20}\mathrm{Ne}$}
\newcommand{\sodios}{$^{23}\mathrm{Na}$}
\newcommand{\silicos}{$^{28}\mathrm{Si}$}
\newcommand{\sulfuros}{$^{32}\mathrm{S}$}
\newcommand{\iroh}{$^{56}\mathrm{Fe}$}
\begin{document}

%%%%%%%%%%%%%%%%%%%%%%%%%%%%%%%%%%%%%%%%
   \title{On the impact of the carbon fusion rate over the properties of superbursts}

   \subtitle{- Numerical simulations of superbursts with MESA -}

%%%%%%%%%%%%%%%%%%%%%%%%%%%%%%%%%%%%%%%%

   \author{M. Nava-Callejas\inst{1}\fnmsep\thanks{Corresponding author: martin.javier.nava.callejas@ulb.be}
        \and S. Goriely, \inst{1}
        \and N. Chamel, \inst{1}
        }

   \institute{Institut d'Astronomie et d'Astrophysique, Université Libre de Bruxelles, 1050 Bruxelles, Belgique}

   \date{Received September 30, 20XX}
 
  \abstract
   {Superbursts are very energetic explosions in the crust of neutron stars in Low-Mass X-ray Binaries (LMXBs). These are triggered by unstable carbon burning at $T\leq 10^{9}$ K. In recent years, there has been a re-examination of the carbon fusion rate, finding that at these temperatures it might be either smaller or higher with respect to the standard rate from Caughlan \& Fowler (1988).}
   {We explore the consequences changing the carbon fusion rate has over the physics of superbursts.}
   {For simulating superbursts, we employ the public code \mesa\ v.24.08.1, as well as four versions of the carbon fusion reaction rate.}
   {A significant enhancement of the reaction rate at $T\leq 10^9$~K would reduce the recurrence and decay times of the superburst, as well as the column depth at ignition. The opposite behavior is observed when the carbon fusion rate is reduced. The maximum temperature reached during the explosion is also sensitive to a change in the carbon fusion rate, leading to either an enhancement or a reduction in the synthesis of $\alpha$-nuclides. These changes are comparable to the effect of reducing the amount of base heating at the bottom of the envelope.}
   {The modelling of X-ray superburst is sensitive to the adopted carbon fusion rate at $T\leq 10^{9}$ K. Given its role for determining the ignition conditions, this rate needs to be better experimentally constrained to 
   reduce the uncertainty in nuclear physics when modeling the superburst light curve and the nucleosynthesis of the explosion.
   At the same time, multizone simulations of superburst also requires improvement in the description of the initially accreted material and stellar conditions prior to the explosion.}


   \maketitle

   \nolinenumbers

%%%%%%%%%%%%%%%%%%%%%%%%%%%%%%%%%%%%%%%%%%%%%%%%%%%%%%%%%%%%%%%%%%%%%%%%%%%%%%%%%%%%%%%%%%%%%%%%%%%%%%%%%%%%%%%%%%%%%%%%%%%%%%%%%%%%%%%%%%%%%%%%%%%%%%%%%%%%%%%%%%%%%%%%%%%%%%%%%%%%%%%%%%%%%%%%%%%%%%%%%%%%%%%%%%%%%%%%%%%%%%%%%%%
%%%%%%%%%%%%%%%%%%%%%%%%%%%%%%%%%%%%%%%%%%%%%%%%%%%%%%%%%%%%%%%%%%%%%%%%%%%%%%%%%%%%%%%%%%%%%%%%%%%%%%%%%%%%%%%%%%%%%%%%%%%%%%%%%%%%%%%%%%%%%%%%%%%%%%%%%%%%%%%%%%%%%%%%%%%%%%%%%%%%%%%%%%%%%%%%%%%%%%%%%%%%%%%%%%%%%%%%%%%%%%%%%%%
\section{Introduction}
%%%%%%%%%%%%%%%%%%%%%%%%%%%%%%%%%%%%%%%%%%%%%%%%%%%%%%%%%%%%%%%%%%%%%%%%%%%%%%%%%%%%%%%%%%%%%%%%%%%%%%%%%%%%%%%%%%%%%%%%%%%%%%%%%%%%%%%%%%%%%%%%%%%%%%%%%%%%%%%%%%%%%%%%%%%%%%%%%%%%%%%%%%%%%%%%%%%%%%%%%%%%%%%%%%%%%%%%%%%%%%%%%%%

% - - - - - - - - - - - - - - - - - - - - - - - - - - - - - - - - - - - - - - - - - - - - - - - - - - - - - %
Superbursts are rare explosive events occurring in the crust of neutron stars in Low-Mass X-ray Binary systems (LMXBs) and are understood as originating from unstable carbon burning. 
They are referred to as ``super'' because of their large energy release, $\sim 10^{41}$ erg, as well as their extended duration ranging from $\sim 3$ to $\sim 20$~hr \citep{Wijnands_2001, Cumming_2001, Strohmayer_2002a, Kuulkers_2004a}.
These features supersede the characteristics of the more frequently observed type I X-ray bursts for which the explosion is due to unstable nuclear burning of \hydros{} and \helioos{} releasing $\sim 10^{38-39}$ erg in less than typically 1~hr.
Table 1 from \cite{intZand_2017} offers a relevant summary of the main properties of the known superbursts before 2018. For the majority of sources, the inferred accretion rate at which superbursts occur is on the order of $\sim 10^{-9}\, M_{\odot}\, \mathrm{yr}^{-1}$. Not all objects accreting at such rate display more than one superburst, although it is fair to say that this can be due to a lack of further monitoring rather than due to the physical underlying mechanism. The average recurrence time between superbursts for sources with this accretion rate is on the order of $\sim 1$ year. One of the most prolific sources exhibiting superbursts is 4U 1636$-$56 \citep{Wijnands_2001, Strohmayer_2002b, Kuulkers_2004b, Kuulkers_2009}. Although its accretion rate is not known, it presumably lies within the same order of magnitude. 
An intriguing exception to this picture is the GX 17+2 source \citep{intZand_2004}, which has displayed up to four superbursts while accreting at $\sim 2\times 10^{-8}\, M_{\odot}\, \mathrm{yr}^{-1}$ and with a recurrence time between events of $\sim 12$ days.
% - - - - - - - - - - - - - - - - - - - - - - - - - - - - - - - - - - - - - - - - - - - - - - - - - - - - - %

% - - - - - - - - - - - - - - - - - - - - - - - - - - - - - - - - - - - - - - - - - - - - - - - - - - - - - %
Concerning nuclear physics, unstable burning of \hydros, \helioos{} and \carbos{} in the crust of neutron stars at temperatures $T\leq 10^{9}$~K is still affected by significant nuclear uncertainties \citep{Meisel_2018}. 
In the case of X-ray bursts, where the primordial fuel is \hydros, it has been shown that the physical conditions during and after the explosion are suitable for triggering long chains of rapid proton captures, modulated by the $\beta^{+}$ decay of proton-rich isotopes as well as by $(\alpha,p)$ transmutations, all of them collectively referred to as the rapid proton-capture process (or rp-process) \citep{Wallace_1981, Schatz_2001}. Such an extensive process, responsible for synthesizing nuclides beyond the $A = 56$ group at moderate physical conditions ($\rho \sim 10^{6}$~g~cm$^{-3}$ and $T\sim 10^{8.5}$~K), is subject to large nuclear uncertainties since not all reaction rates are experimentally constrained. While some reactions have been identified as playing a more critical role than others during the rp-process \citep{Cyburt_2010, Cyburt_2016}, as for example the CNO-breaking $^{15}$O($\alpha,\gamma$)$^{19}$Ne 
\citep{Fisker_2006, Meisel_2018}, a complete analysis on the impact of rate uncertainties remains a matter of ongoing research.
In contrast, superbursts are, at a very superficial level, a much simpler scenario since the lifetime of \carbos{}, in the absence of \hydros{} and \helioos{} in the mixture of ashes, is mainly governed by its fusion rate, responsible for both increasing the mass fraction of \neos{} and \sodios{} and releasing $\alpha$ and $p$ particles which can then be re-captured by other nuclides in the mixture.
% 3. Simulations of superbursts
So far  and to the best of our knowledge, multi-zone simulations of superbursts combining hydrodynamics and a large network of reactions have only been carried out with the private stellar evolution code \kepler, as reported in \cite{KH2011, KH2012}, hereafter KH11 and KH12 respectively\footnote{From a historical point of view, however, it is worth mentioning that these papers are consistent with early attempts reported by independent groups, as in \cite{Weinberg_2006, Weinberg_2007, Noel_2008}.}.
In KH11, it was explored how characteristics of superbursts, such as the recurrence time or the column depth of explosion, were affected by changes in the mass accretion rate and the base luminosity of the envelope. KH12 aimed at producing a more realistic physical scenario, displaying standard X-ray bursts before and after the superburst. This was achieved by superimposing \hydros{} and \helioos{} layers on top of a layer with sufficient carbon to trigger an explosion. The results of these studies have been highly influential since the properties of their simulated superbursts are in relative good agreement with those inferred from observations. For instance, the luminosity decay time (defined as the time it takes for the luminosity to fall below its maximum value by a factor $\exp(1)$), which is in the order of hours to $\sim$ day. A second success of these models is in predicting recurrence times between $\sim 10$ days and $\sim 10$ years, depending on the assumed mass accretion rate. However, it should be stressed that the fit of a single superburst light curve with such simulation cannot unambiguously determine the mass accretion rate and the luminosity at the base of the envelope, these two parameters being degenerate.
% - - - - - - - - - - - - - - - - - - - - - - - - - - - - - - - - - - - - - - - - - - - - - - - - - - - - - %

% - - - - - - - - - - - - - - - - - - - - - - - - - - - - - - - - - - - - - - - - - - - - - - - - - - - - - %
Although inaccessible from observations, the chemical composition of the superburst ashes represents another important contribution from KH11 and KH12. According to these simulations, the distribution of ashes exhibits two peaks around \iroh{} and \silicos{} respectively. While such result has been taken as the canonical ``composition of superburst ashes'' elsewhere \citep[e.g.][]{Lau_2018, Shchechilin_2021, Shchechilin_2022, Jain_2025}, the fact that they resulted from simulations accreting $20\%$ \carbos{} and 80\% \iroh{} rises the question of the sensitivity of the 
distribution of nuclides after a superburst to the initial composition.
In addition, we can also ponder if the mass accretion rate plays a role in enhancing or not the synthesis of high-$Z$ nuclides, similarly to what happens with the rp-process for \hydros/\helioos{} burning (e.g. \cite{Schatz_1999, Schatz_2001}). 
% - - - - - - - - - - - - - - - - - - - - - - - - - - - - - - - - - - - - - - - - - - - - - - - - - - - - - %

% - - - - - - - - - - - - - - - - - - - - - - - - - - - - - - - - - - - - - - - - - - - - - - - - - - - - - %
From the point of view of nuclear physics, the scope of these works has room for further explorations. In first place, both KH11 and KH12 assume that the superburts can be triggered if $20\%$ of the fuel mass is composed of \carbos. This amount is based on the theoretical calculations from \cite{Cumming_2004, Cumming_2006}. In these latter works, it was shown that the ignition column depth $y_{\mathrm{ignition}}$ and the required mass fraction of \carbos{} for triggering superbursts can be constrained from the luminosity curve. In particular, when applying their model to the superbursts from KS 1731$-$260 and 4U 1636$-$536 they showed that $y_{\mathrm{ignition}}\approx 10^{11-12}$ g cm$^{-2}$ and the required amount of \carbos{} is $\sim 20\%$ of the total fuel mass. The origin of such a high amount of \carbos{} is still a matter of debate. For a solar-like composition of accreted matter, i.e. $\sim 70\%$ of \hydros{} and $\sim 28\%$ of \helioos{}, the unstable nuclear burning results in mass fractions for \carbos{} which are typically $\la  10\%$ of the total composition of the ashes \citep{Schatz_2001, Meisel_2019}. On the other hand, the stable burning of the same composition actually allows to reach mass fractions of \carbos{} close to the required $20\%$ \citep{Stevens_2014, Zamfir_2014}, in particular for a narrow range of mass accretion rates around $10\%$ of the Eddington accretion rate, typical of superbursts \citep[e.g.][]{KH_2016}.
However, the stabilization requires a mass accretion rate that is usually higher than the one inferred from observations \citep{Galloway_2021}.
% - - - - - - - - - - - - - - - - - - - - - - - - - - - - - - - - - - - - - - - - - - - - - - - - - - - - - %

% - - - - - - - - - - - - - - - - - - - - - - - - - - - - - - - - - - - - - - - - - - - - - - - - - - - - - %
A second aspect to be considered is the carbon fusion rate. KH11 and KH12 adopted the rate from \cite{CF88_bib} - hereafter CF88. However, over the last 20 years there has been a substantial amount of work attempting to constrain experimentally the carbon fusion rate at temperatures relevant to explosive C-burning. On the one hand, \cite{Pignatari_2013} showed that the fractions of the total carbon fusion rate going into the $p$- and $\alpha-$ channels should be updated from $\sim 44\%$ and $\sim 56\%$ \citep{CF88_bib} into $\sim 35\%$ and $65\%$, respectively. On the other hand, \cite{Monpribat22} argued that the non-resonant carbon fusion rate at $T\leq 10^{9}$~K - temperatures prevailing in neutron-star envelopes both during standard X-ray bursts and superbursts- might be $\sim 10^{-3}$ times smaller than predicted by CF88. In contrast, \cite{Tumino18} claimed that the low-energy resonances might actually increase the rate by a factor of typically $10^{3}$ in the same temperature interval.
% - - - - - - - - - - - - - - - - - - - - - - - - - - - - - - - - - - - - - - - - - - - - - - - - - - - - - %

% - - - - - - - - - - - - - - - - - - - - - - - - - - - - - - - - - - - - - - - - - - - - - - - - - - - - - %
The purpose of the present work is to explore the impact the carbon fusion reaction rate has on the properties of superbursts, namely, the peak luminosity, the decay and recurrence times, as well as the composition of the ashes based in realistic astrophysical simulations.
In Sect.~\ref{section:methods} we describe the software employed as well as the physical parameters of our simulations. In Sect.~\ref{sec:results} we present the results for different mass accretion rates. Finally, in Sect.~\ref{sec:ending} we discuss the implications of our findings and prospects for future works.
% - - - - - - - - - - - - - - - - - - - - - - - - - - - - - - - - - - - - - - - - - - - - - - - - - - - - - %

%%%%%%%%%%%%%%%%%%%%%%%%%%%%%%%%%%%%%%%%%%%%%%%%%%%%%%%%%%%%%%%%%%%%%%%%%%%%%%%%%%%%%%%%%%%%%%%%%%%%%%%%%%%%%%%%%%%%%%%%%%%%%%%%%%%%%%%%%%%%%%%%%%%%%%%%%%%%%%%%%%%%%%%%%%%%%%%%%%%%%%%%%%%%%%%%%%%%%%%%%%%%%%%%%%%%%%%%%%%%%%%%%%%
%%%%%%%%%%%%%%%%%%%%%%%%%%%%%%%%%%%%%%%%%%%%%%%%%%%%%%%%%%%%%%%%%%%%%%%%%%%%%%%%%%%%%%%%%%%%%%%%%%%%%%%%%%%%%%%%%%%%%%%%%%%%%%%%%%%%%%%%%%%%%%%%%%%%%%%%%%%%%%%%%%%%%%%%%%%%%%%%%%%%%%%%%%%%%%%%%%%%%%%%%%%%%%%%%%%%%%%%%%%%%%%%%%%

%%%%%%%%%%%%%%%%%%%%%%%%%%%%%%%%%%%%%%%%%%%%%%%%%%%%%%%%%%%%%%%%%%%%%%%%%%%%%%%%%%%%%%%%%%%%%%%%%%%%%%%%%%%%%%%%%%%%%%%%%%%%%%%%%%%%%%%%%%%%%%%%%%%%%%%%%%%%%%%%%%%%%%%%%%%%%%%%%%%%%%%%%%%%%%%%%%%%%%%%%%%%%%%%%%%%%%%%%%%%%%%%%%%
%%%%%%%%%%%%%%%%%%%%%%%%%%%%%%%%%%%%%%%%%%%%%%%%%%%%%%%%%%%%%%%%%%%%%%%%%%%%%%%%%%%%%%%%%%%%%%%%%%%%%%%%%%%%%%%%%%%%%%%%%%%%%%%%%%%%%%%%%%%%%%%%%%%%%%%%%%%%%%%%%%%%%%%%%%%%%%%%%%%%%%%%%%%%%%%%%%%%%%%%%%%%%%%%%%%%%%%%%%%%%%%%%%%
\section{Methodology}\label{section:methods}
%%%%%%%%%%%%%%%%%%%%%%%%%%%%%%%%%%%%%%%%%%%%%%%%%%%%%%%%%%%%%%%%%%%%%%%%%%%%%%%%%%%%%%%%%%%%%%%%%%%%%%%%%%%%%%%%%%%%%%%%%%%%%%%%%%%%%%%%%%%%%%%%%%%%%%%%%%%%%%%%%%%%%%%%%%%%%%%%%%%%%%%%%%%%%%%%%%%%%%%%%%%%%%%%%%%%%%%%%%%%%%%%%%%

% - - - - - - - - - - - - - - - - - - - - - - - - - - - - - - - - - - - - - - - - - - - - - - - - - - - - - %
For simulating the evolution of neutron star envelopes undergoing mass accretion, we employed the publicly available code \mesa{} v.24.08.1 \citep{Jermyn_2023aa, Paxton_2019aa, Paxton_2018aa, Paxton_2015aa, Paxton_2013aa, Paxton_2011aa}. This code has proven to be reliable in the X-ray burst simulation of the thermonuclear explosions from unstable H/He burning, in particular when compared against the results from the private code \kepler \citep{Meisel_2018, Johnston_2020}. Similarly to \cite{Nava-Callejas_2025, Cavecchi_2026}, we replaced the tabulated radiative opacities by the prescription from Bill Wolf's\footnote{\url{https://billwolf.space/projects/leiden_2019/}}. This prescription consists in combining the analytic fits from \cite{Schatz_1999} with the recent electron scattering results from \cite{Poutanen_2017} and applying a corrective factor from \cite{Potekhin_2001}. General relativistic corrections were not taken into account for the present simulations since \mesa{} requires additional modifications to the radiative transport to ensure a self-consistent treatment (see Appendix A from \citealp{Nava-Callejas_2025}). Convection and thermohaline were included in our simulations following the Mixing Length Theory+ scheme \citep{Paxton_2013aa}, which considers solar-like values for the mixing length and the thermohaline coefficient. To prevent an artificial expansion of the low-density layers of the envelope - a phenomenon also observed in the numerical simulations with \kepler{} \citep{KH2011} - we adopted the boundary condition of fixed pressure setting the surface at $P_{\mathrm{surf}} = 10^{20}$ erg cm$^{-3}$. As is standard in the field, and unless otherwise stated, the initial chemical composition of the envelope is of pure \iroh.
% - - - - - - - - - - - - - - - - - - - - - - - - - - - - - - - - - - - - - - - - - - - - - - - - - - - - - %

% - - - - - - - - - - - - - - - - - - - - - - - - - - - - - - - - - - - - - - - - - - - - - - - - - - - - - %
Regarding the chemical composition of the accreted material, for our exploration of superbursts in \mesa\ we took the standard values in the literature required to simulate superbursts \citep{Cumming_2006, KH2011}, i.e. $20\%$ \carbos{} and $80\%$ \iroh. For a similar reason, we opted for simulating superburts primordially with a reduced network of reactions, \approxai, which includes all $A \leq 56$ nuclides in and above the valley of stability. We considered this as a good starting point for modelling superbursts since, as shown by KH11 and KH12, the unstable carbon burning is not likely to synthesize material beyond $A = 64$. To ensure this is the case we also ran a few simulations with a slightly increased network of reactions, \approxaii, with $A\leq 64$ nuclides in and above the valley of stability.
% - - - - - - - - - - - - - - - - - - - - - - - - - - - - - - - - - - - - - - - - - - - - - - - - - - - - - %

% - - - - - - - - - - - - - - - - - - - - - - - - - - - - - - - - - - - - - - - - - - - - - - - - - - - - - %
Considering the observational constraints from \cite{Kuulkers_2004a, intZand_2017}, as well as the numerical simulations of KH11 and KH12, we selected two values for the mass accretion rate: \middmdot $= 5.88\times 10^{-9}\, M_{\odot}\, \mathrm{yr}^{-1}$ and \highmdot $= 2\times 10^{-8}\, M_{\odot}\, \mathrm{yr}^{-1}$. These values roughly correspond to $\approx 0.3\dot{M}_{\mathrm{Edd}}$ and $\approx \dot{M}_{\mathrm{Edd}}$ respectively, with $\dot{M}_\mathrm{Edd} = 1.75\times 10^{-8}\ M_{\odot}\, \mathrm{yr}^{-1}$ the Eddington mass accretion rate.
\mesa, similarly to \kepler, considers a fixed luminosity at the deepest layer of the envelope as boundary condition. The base luminosity $L_b$ of the envelope for each case was chosen following the reported values in KH11 that lead to an unstable carbon burning: for $\dot{M}_{\mathrm{moderate}}$ we set $L_{b} = 25L_{\odot}$ while for $\dot{M}_{\mathrm{high}}$ we took $L_{b} = 50L_{\odot}$. 
KH11 report the base luminosity of the envelope as $Q_b$, in units of MeV per baryon. Our numerical values are equivalent to 0.26 and 0.16 MeV per baryon, respectively, according to the conversion rule between $Q_b$ and $L_b$ given by $L_{b} = Q_{b} \times \dot{M}/m_{u}$, with $m_{u}$ the atomic mass unit.
% - - - - - - - - - - - - - - - - - - - - - - - - - - - - - - - - - - - - - - - - - - - - - - - - - - - - - %

\begin{figure}[ht!]
    \centering
    \includegraphics[width=1\hsize]{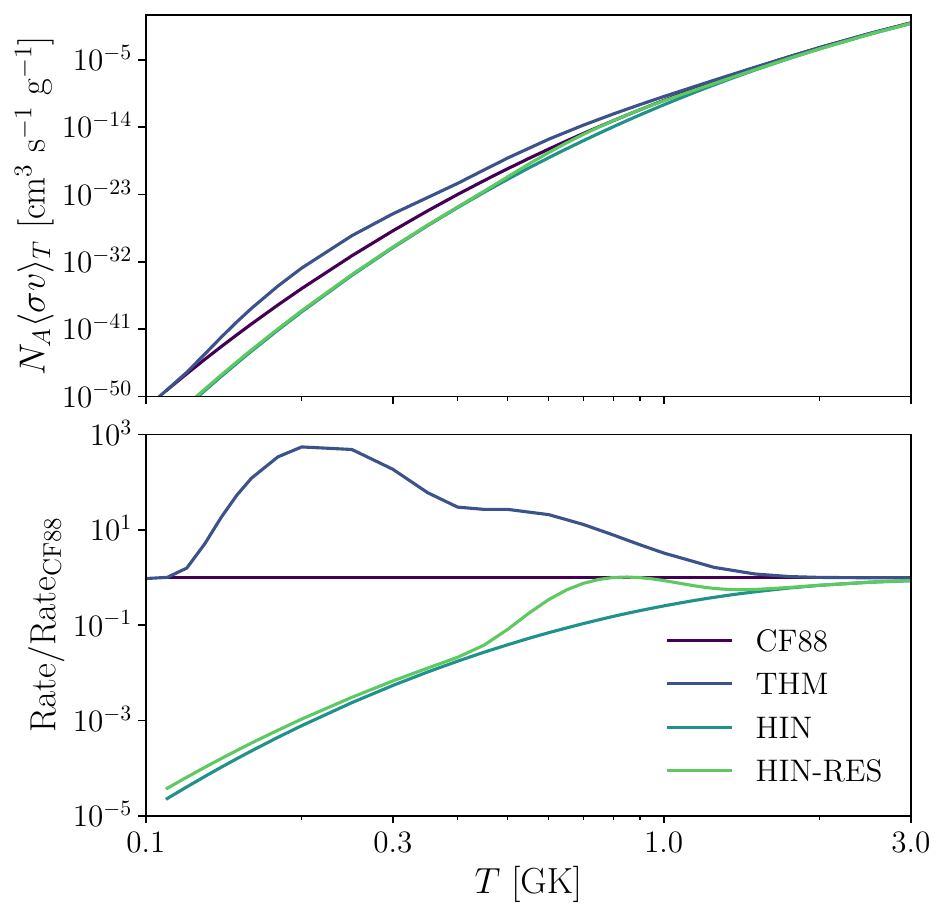}
    \caption{Top panel: different versions of the total \carbos{} fusion reaction rate as a function of temperature considered in the present work. Bottom panel: reaction rate ratio with respect to CF88.}
    \label{fig:reaction_rates_comparions}
\end{figure}

% - - - - - - - - - - - - - - - - - - - - - - - - - - - - - - - - - - - - - - - - - - - - - - - - - - - - - %
In the present work, we compare four versions of the carbon fusion reaction rate. The first of them is the classical reaction rate CF88. Since it is the default one in \mesa, as well as in former astrophysical studies, we take it as our fiducial simulation. From the total reaction rate, $\approx 44\%$ goes into the $p$ channel while the remainder $56\%$ goes into the $\alpha$-channel.
Two alternatives for the carbon fusion rate come from \cite{Monpribat22} and correspond to the HIN and HIN-RES models, which are in relatively good agreement with the measurements obtained by the STELLA collaboration \citep{Monpribat22}. In the HIN model, the cross section of the carbon-carbon fusion follows the empirical hindrance model. The HIN-RES model additionally includes the low-lying resonance proposed by \cite{Spillane_2007}, although the existence of such a resonance remains uncertain. In contrast to the CF88 rate, but following a recent prescription from experimental data \citep{Pignatari_2013}, the fraction of the total rate going into the $p$- and $\alpha$- channels for both HIN and HIN-RES rates are set to $35\%$ and $65\%$, respectively.
The fourth version of the carbon fusion rate we consider comes from the measurements performed at the Laboratory for Underground Nuclear Astrophysics (LUNA) using the Trojan Horse method \citep{Tumino18}. Labeled as THM, we employ the same updated ratios as in HIN and HIN-RES, i.e. $35\%$ of the total rate into the $p$-channel and $65\%$ into the $\alpha$-channel. 
The four total reaction rates are compared in the top panel of Fig.~\ref{fig:reaction_rates_comparions}, while in the bottom panel we show the ratio of all displayed reaction rates with respect to CF88. While the four reaction rates are in good agreement at $T\geq 2\times 10^{9}$ K, at lower temperatures they exhibit quite different behaviors. More specifically, at $T \sim 2-3\times10^8$~K, the THM carbon fusion rate is about  $10^3$ times higher than CF88 while the HIN/HIN-RES one is about $10^3$ times smaller.
The $n-$channel of the carbon fusion rate, while present in our numerical simulations, is of reduced importance for superbursts in comparison to the $p-$ and $\alpha-$ channels due to its endothermic nature and thus it has not been modified in this first exploration. While the updated branch ratio from \cite{Bucher_2015} shows its importance for neutron processes at $T\gg 2\times 10^{9}$~K we see that the discrepancy with respect to the CF88 rate is of a factor of $3$. For the sake of clarity, in Appendix~\ref{appendix:n_channel_supeburst} we show that the effects of this rate over the characteristics of superbursts is of second order and does not modify our main findings.
% - - - - - - - - - - - - - - - - - - - - - - - - - - - - - - - - - - - - - - - - - - - - - - - - - - - - - %

%----------------------------------------------------------------------------------------------------------------------------------%
%----------------------------------------------------------------------------------------------------------------------------------%

%%%%%%%%%%%%%%%%%%%%%%%%%%%%%%%%%%%%%%%%%%%%%%%%%%%%%%%%%%%%%%%%%%%%%%%%%%%%%%%%%%%%%%%%%%%%%%%%%%%%%%%%%%%%%%%%%%%%%%%%%%%%%%%%%%%%%%%%%%%%%%%%%%%%%%%%%%%%%%%%%%%%%%%%%%%%%%%%%%%%%%%%%%%%%%%%%%%%%%%%%%%%%%%%%%%%%%%%%%%%%%%%%%%
%%%%%%%%%%%%%%%%%%%%%%%%%%%%%%%%%%%%%%%%%%%%%%%%%%%%%%%%%%%%%%%%%%%%%%%%%%%%%%%%%%%%%%%%%%%%%%%%%%%%%%%%%%%%%%%%%%%%%%%%%%%%%%%%%%%%%%%%%%%%%%%%%%%%%%%%%%%%%%%%%%%%%%%%%%%%%%%%%%%%%%%%%%%%%%%%%%%%%%%%%%%%%%%%%%%%%%%%%%%%%%%%%%%

%%%%%%%%%%%%%%%%%%%%%%%%%%%%%%%%%%%%%%%%%%%%%%%%%%%%%%%%%%%%%%%%%%%%%%%%%%%%%%%%%%%%%%%%%%%%%%%%%%%%%%%%%%%%%%%%%%%%%%%%%%%%%%%%%%%%%%%%%%%%%%%%%%%%%%%%%%%%%%%%%%%%%%%%%%%%%%%%%%%%%%%%%%%%%%%%%%%%%%%%%%%%%%%%%%%%%%%%%%%%%%%%%%%
%%%%%%%%%%%%%%%%%%%%%%%%%%%%%%%%%%%%%%%%%%%%%%%%%%%%%%%%%%%%%%%%%%%%%%%%%%%%%%%%%%%%%%%%%%%%%%%%%%%%%%%%%%%%%%%%%%%%%%%%%%%%%%%%%%%%%%%%%%%%%%%%%%%%%%%%%%%%%%%%%%%%%%%%%%%%%%%%%%%%%%%%%%%%%%%%%%%%%%%%%%%%%%%%%%%%%%%%%%%%%%%%%%%
\section{Results}\label{sec:results}
%%%%%%%%%%%%%%%%%%%%%%%%%%%%%%%%%%%%%%%%%%%%%%%%%%%%%%%%%%%%%%%%%%%%%%%%%%%%%%%%%%%%%%%%%%%%%%%%%%%%%%%%%%%%%%%%%%%%%%%%%%%%%%%%%%%%%%%%%%%%%%%%%%%%%%%%%%%%%%%%%%%%%%%%%%%%%%%%%%%%%%%%%%%%%%%%%%%%%%%%%%%%%%%%%%%%%%%%%%%%%%%%%%%

%%--%%--%%--%%--%%--%%--%%--%%--%%--%%--%%--%%--%%--%%--%%--%%--%%--%%--%%--%%--%%--%%--%%--%%--%%--%%--%%--%%--%%--%%--%%--%%--%%--%%--%%--%%--%%--%%--%%--%%--%%--%%--%%--%%--%%--%%--%%--%%--%%--%%--%%--%%--%%--%%--%%--%%--%%
%%--%%--%%--%%--%%--%%--%%--%%--%%--%%--%%--%%--%%--%%--%%--%%--%%--%%--%%--%%--%%--%%--%%--%%--%%--%%--%%--%%--%%--%%--%%--%%--%%--%%--%%--%%--%%--%%--%%--%%--%%--%%--%%--%%--%%--%%--%%--%%--%%--%%--%%--%%--%%--%%--%%--%%--%%
\subsection{Simulations with the network \approxai}\label{subsec:approx_a56}
%%--%%--%%--%%--%%--%%--%%--%%--%%--%%--%%--%%--%%--%%--%%--%%--%%--%%--%%--%%--%%--%%--%%--%%--%%--%%--%%--%%--%%--%%--%%--%%--%%--%%--%%--%%--%%--%%--%%--%%--%%--%%--%%--%%--%%--%%--%%--%%--%%--%%--%%--%%--%%--%%--%%--%%--%%

%----------------------------------------------------------------------------------------------------------------------------------------------------------------------------------------------------------------------------%
\begin{table*}[ht!]
\caption{Properties of superburst with the network \approxai}\label{tab:table_1_superbursts_parameters_a56}
\centering
\begin{tabular}{c|ccc|ccc}
\hline
\hline
& & \middmdot & & & \highmdot & \\
Reaction rate & $t_{\mathrm{recurrence}}$ & $t_{\mathrm{decay}}$ & $y_{\mathrm{ignition}}$ & $t_{\mathrm{recurrence}}$ & $t_{\mathrm{decay}}$ & $y_{\mathrm{ignition}}$\\
  & [day] & [hr]  & [$10^{11}\, \mathrm{g\ cm}^{-2}$] & [day] & [hr]  & [$10^{11}\, \mathrm{g\ cm}^{-2}$]\\
\hline
CF88 & 241.5 & 2.2 & 4.20 & 16.4 & 0.40 & 0.96\\
THM & 100.2 & 1.2 & 1.90 & 10.8 & 0.29 & 0.62\\
HIN & 364.3  & 3.5 & 7.96 & 20.7 & 0.62 & 1.36\\
HIN-RES & 187.2 & 2.8 & 4.18 & 16.3 & 0.42 & 0.96\\
\hline
\hline
\end{tabular}
\end{table*}
%----------------------------------------------------------------------------------------------------------------------------------------------------------------------------------------------------------------------------%

% - - - - - - - - - - - - - - - - - - - - - - - - - - - - - - - - - - - - - - - - - - - - - - - - - - - - - %
In Table~\ref{tab:table_1_superbursts_parameters_a56}, we summarize the characteristics of the superbursts for the two accretion rates \middmdot and \highmdot, according to the adopted carbon fusion reaction rate. These include the recurrence time between 
superbursts, the decay time and the maximum column depth $y_{\mathrm{ignition}}$ reached by the accreted mass before the explosion.
As frequently assumed, we consider that KH11 adopted the CF88 version of the carbon fusion rate.
We thus first compare the results from our CF88 simulations with the values reported in their paper. 
Overall, we find a good agreement. For example, at \highmdot, the recurrence time is in the range of $\sim 10^{-2}$ years $\approx 10$ days and $y_{\mathrm{ignition}}$ is in both cases $\geq 6\times 10^{10}$ g cm$^{-2}$.
Let us now compare the results from our simulations with different versions of the carbon fusion rate. At both \middmdot and \highmdot we observe a reduction or an enhancement of both timescales and the maximum column depth depending on whether the carbon fusion rate is increased or decreased at $T\la 10^{9}$ K. For both mass accretion rates, and with respect to the CF88-reaction rate, the simulations with the THM data lead to a short recurrence time between superbursts. On the other hand, the recurrence time is increased  when use is made of the HIN rate. With respect to CF88, the location of the ignition point is also shifted towards a less dense environment in the THM case and to a more dense one in the HIN case.
% - - - - - - - - - - - - - - - - - - - - - - - - - - - - - - - - - - - - - - - - - - - - - - - - - - - - - %

%----------------------------------------------------------------------------------------------------------------------------------------------------------------------------------------------------------------------------%
\begin{figure*}[ht!]
    \sidecaption
    \includegraphics[width=0.35\textwidth]{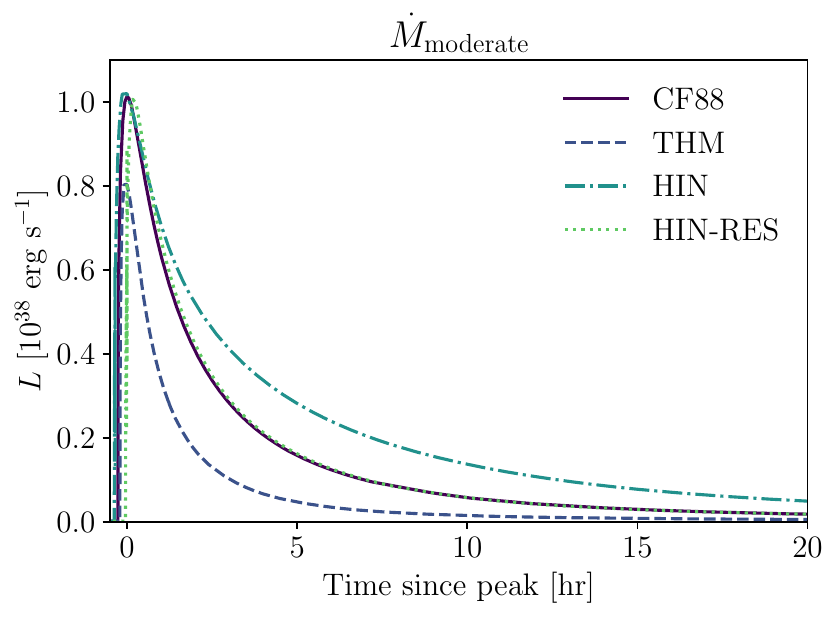}
    \includegraphics[width=0.35\textwidth]{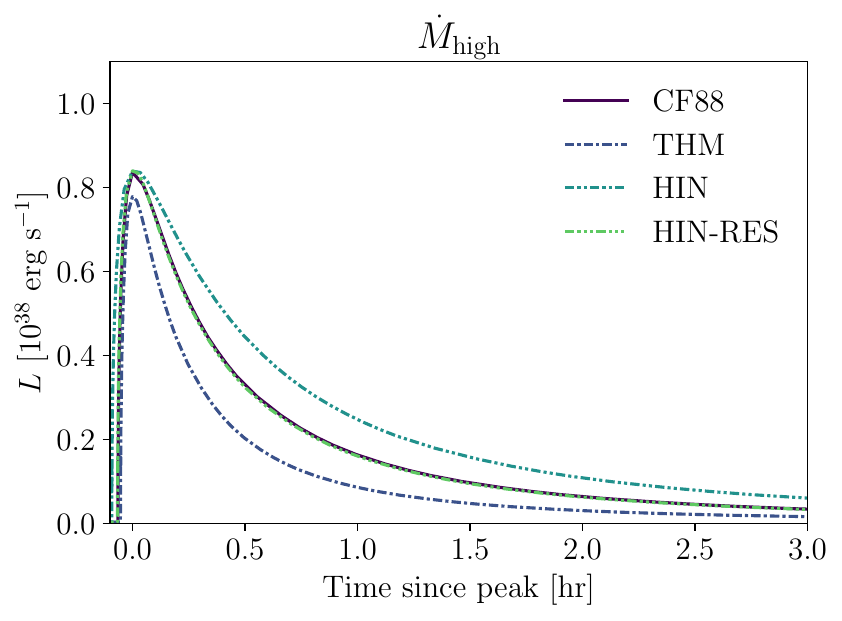}
    \caption{Superburst luminosity curve for each version of the carbon fusion rate considered in this work. Time has been shifted in each case in order to ensure the luminosity peaks occurs at $t=0$ hr. Left panel: simulations at \middmdot. Right panel: simulations at \highmdot.}
    \label{fig:luminosity_profiles_1}
\end{figure*}
%----------------------------------------------------------------------------------------------------------------------------------------------------------------------------------------------------------------------------%

%----------------------------------------------------------------------------------------------------------------------------------------------------------------------------------------------------------------------------%
\begin{figure*}[ht!]
    \sidecaption
    \includegraphics[width=0.35\textwidth]{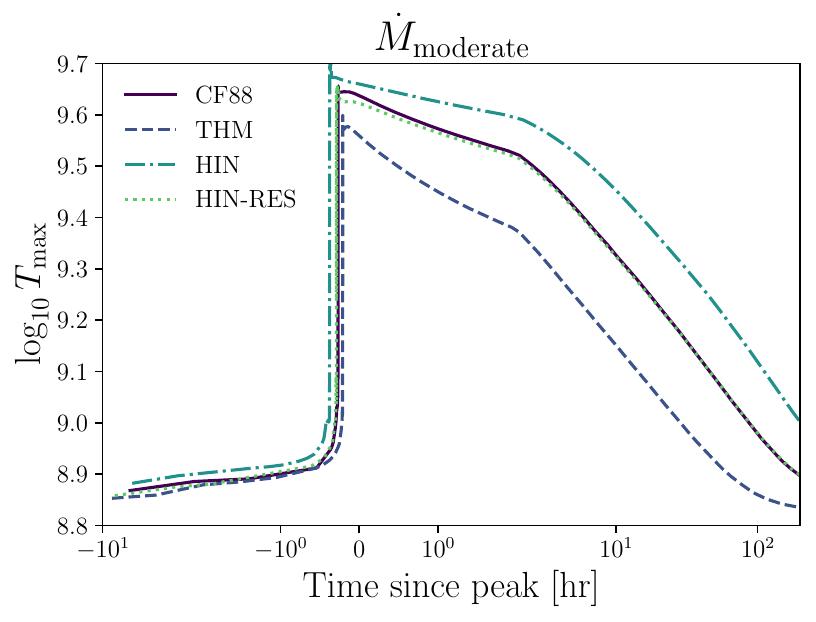}
    \includegraphics[width=0.35\textwidth]{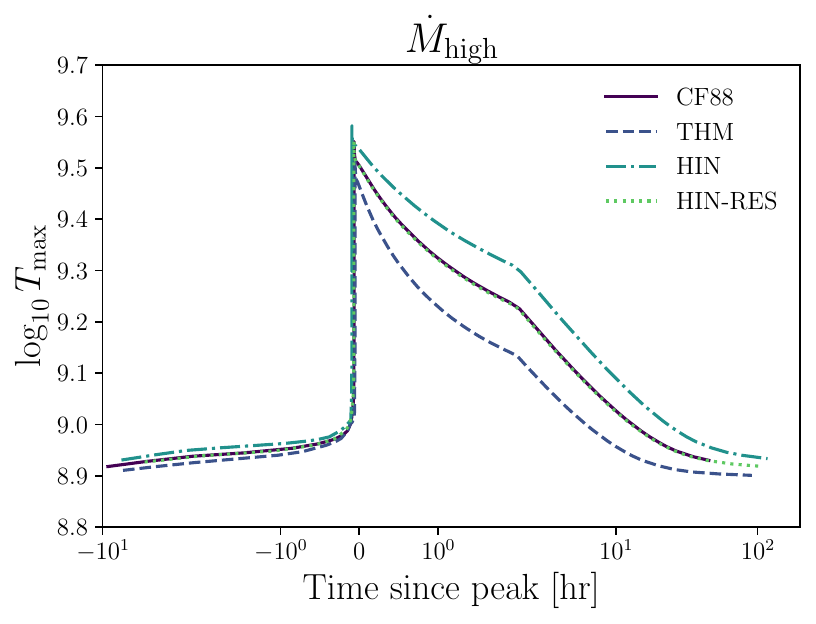}
    \caption{Maximum temperature of the envelope (in K) as a function of time (in hours) for each version of the carbon fusion rate. Time has been shifted to ensure that the luminosity peak occurs at $t = 0$. Left panel: simulations at \middmdot. Right panel: simulations at \highmdot.}
    \label{fig:maximum_temperatures_1}
\end{figure*}
%----------------------------------------------------------------------------------------------------------------------------------------------------------------------------------------------------------------------------%

%----------------------------------------------------------------------------------------------------------------------------------------------------------------------------------------------------------------------------%
\begin{figure*}[ht!]
    \centering
    \includegraphics[width=0.45\textwidth]{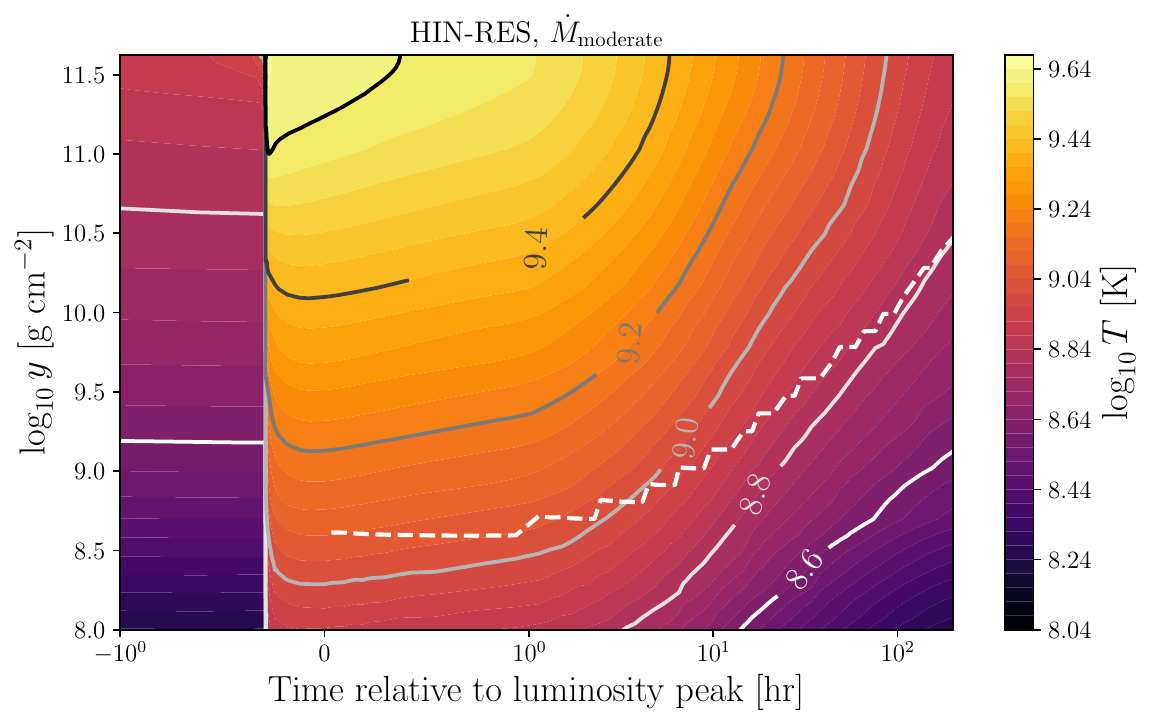}
    \includegraphics[width=0.45\textwidth]{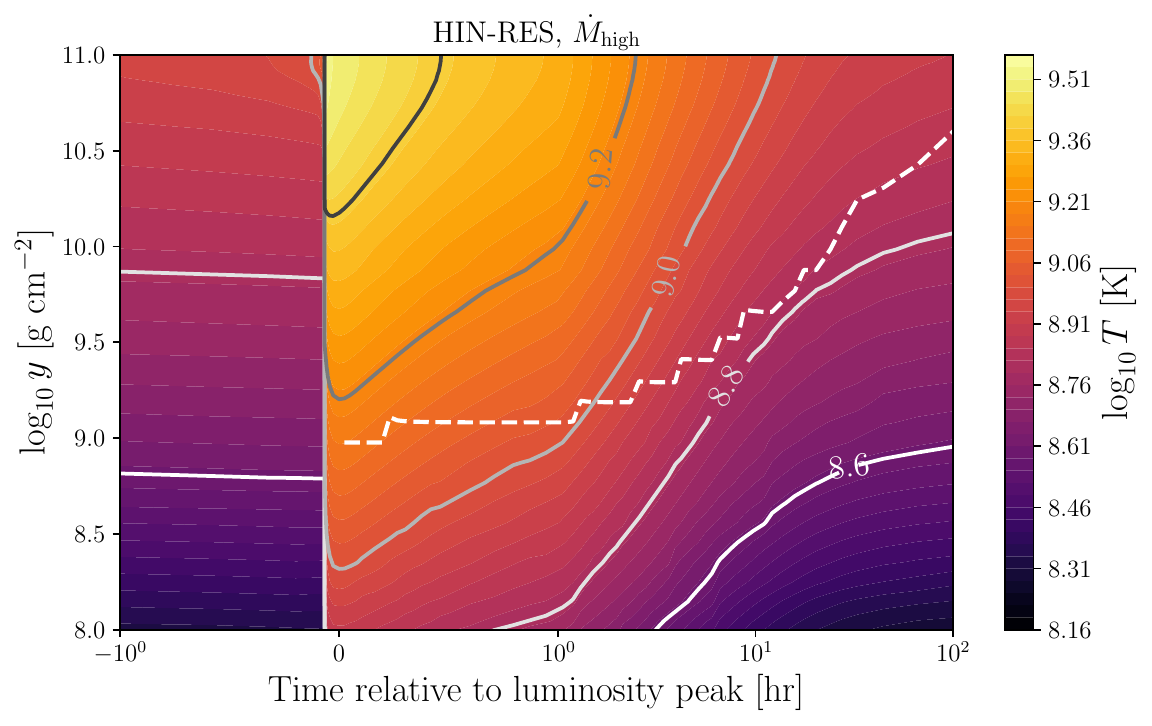}
    \caption{Contour plot for the temperature of the envelope and crust as a function of time and column depth for the HIN-RES simulations. Time has been shifted to ensure that the luminosity peak occurs at $t = 0$. The dashed contour indicates the instantaneous column depth at $t>0$ for which the mass fraction of \carbos{} is equal to $10^{-6}$. Left panel: simulation at \middmdot. Right panel: simulation at \highmdot.}
    \label{fig:maximum_temperatures_2}
\end{figure*}
%----------------------------------------------------------------------------------------------------------------------------------------------------------------------------------------------------------------------------%

%----------------------------------------------------------------------------------------------------------------------------------------------------------------------------------------------------------------------------%
\begin{figure*}[ht!]
    \sidecaption
    \includegraphics[width=0.7\textwidth]{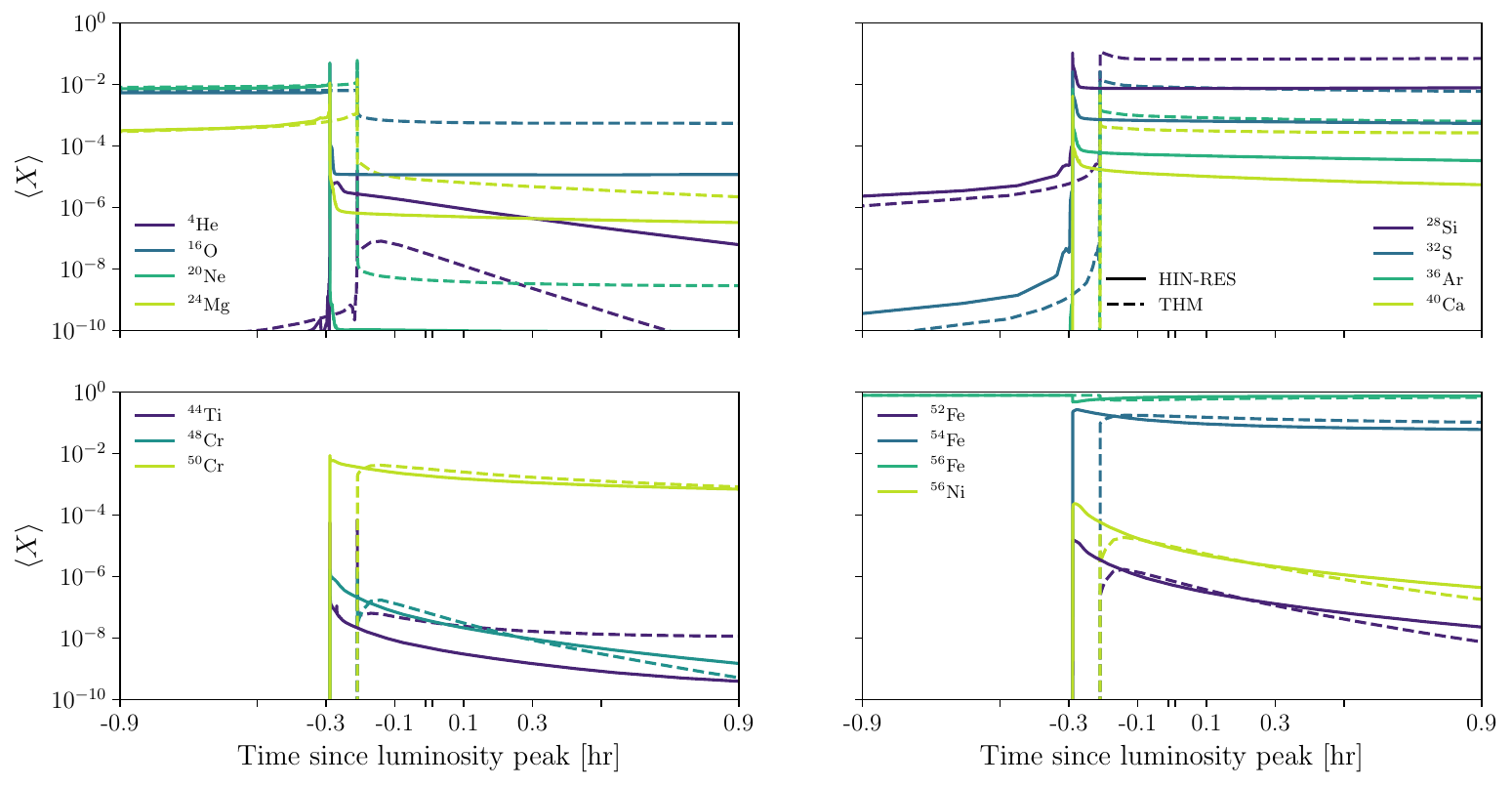}
    \caption{Time evolution of the average mass fractions of various nuclear species at \middmdot\ for the THM (dashed lines) and HIN-RES solid lines) simulations during a superburst. See main text for further details.}
    \label{fig:mass_fractions_1}
\end{figure*}
%----------------------------------------------------------------------------------------------------------------------------------------------------------------------------------------------------------------------------%

%----------------------------------------------------------------------------------------------------------------------------------------------------------------------------------------------------------------------------%
\begin{figure}[ht!]
    \centering
    \includegraphics[width=0.45\textwidth]{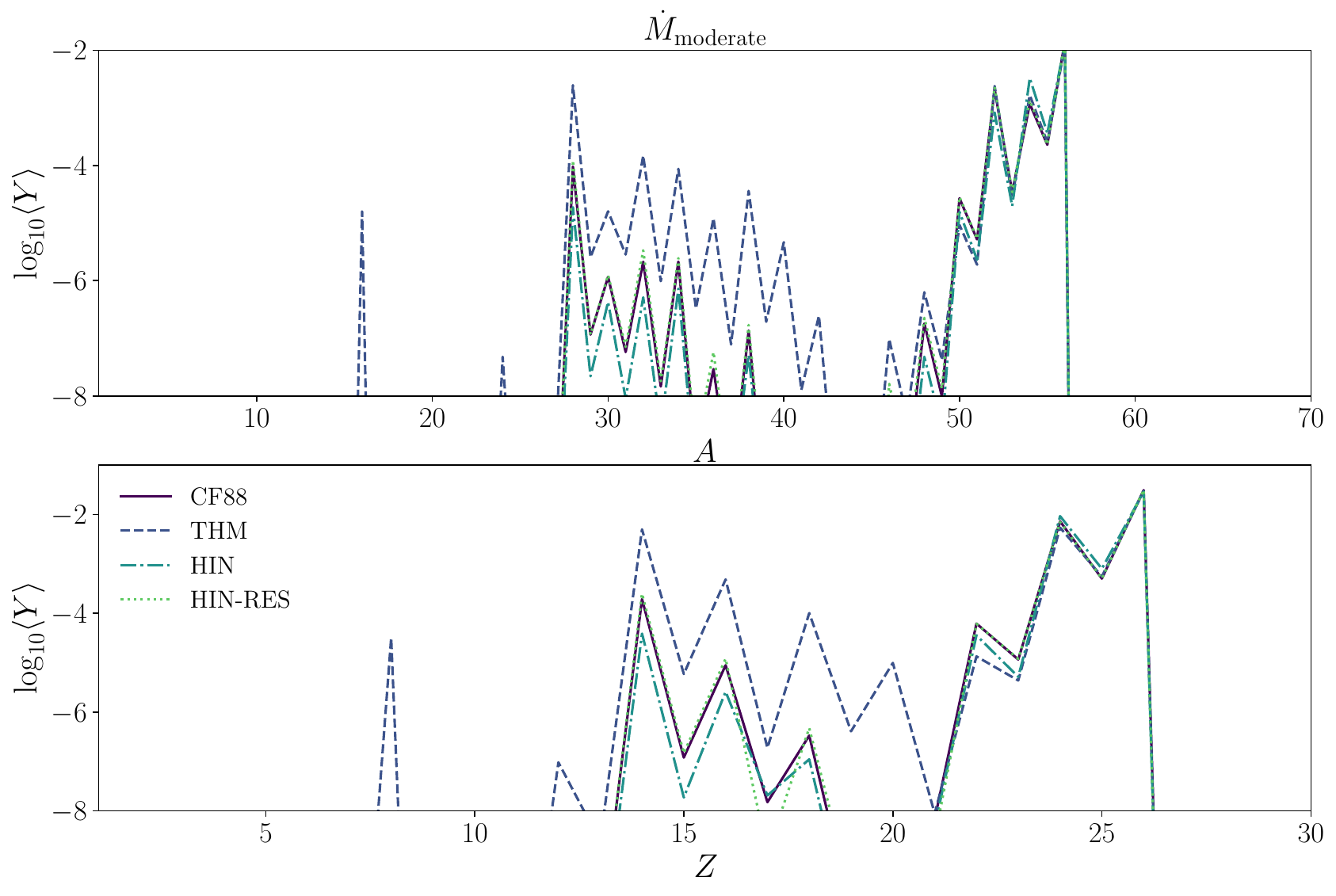}
    \includegraphics[width=0.45\textwidth]{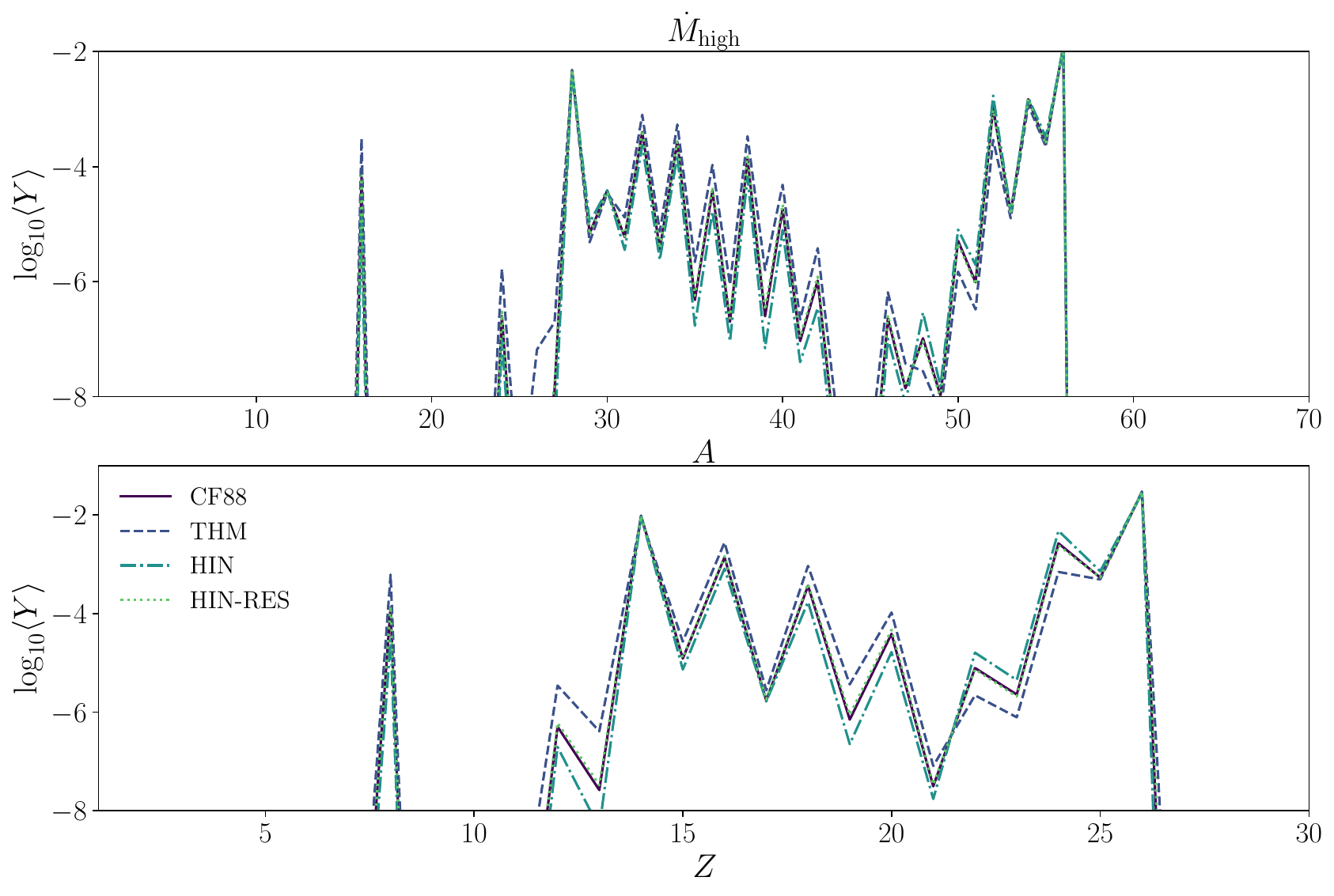}
    \caption{Distribution of average abundances $\langle Y \rangle$ as a function of mass number $A$ - upper panels - and as a function of charge number $Z$ - lower panels - for the ashes of the first superburst in the case of \middmdot - upper figure - and of the second superburst in the case of \highmdot - lower figure. The labels correspond to the four carbon fusion rates adopted in this work. See main text for further details.}
    \label{fig:distribution_ashess_1}
\end{figure}
%----------------------------------------------------------------------------------------------------------------------------------------------------------------------------------------------------------------------------%

% - - - - - - - - - - - - - - - - - - - - - - - - - - - - - - - - - - - - - - - - - - - - - - - - - - - - - %
Figure~\ref{fig:luminosity_profiles_1} compares the superburst luminosity curves at \middmdot and \highmdot obtained with the four carbon fusion reaction rates. To perform a fair comparison, we have shifted the time coordinate of each model to ensure the location of the maximum in luminosity occurs at $t=0$. The first characteristic to notice is that in both panels the CF88 and HIN-RES simulations are very similar in peak luminosity - around $10^{38}$ erg s$^{-1}$ at \middmdot and $8\times 10^{37}$ erg s$^{-1}$ at \highmdot - as well as in the decay profile, having an overall decay time of $\sim 2.5$ hr at \middmdot and $\sim 0.40$ hr at \highmdot. For both accretion rates, the CF88 and HIN-RES simulations represent a middle point with respect to the other two simulations, which display either a shortened decay time, in the case of the THM rate, or an increased decay time in the case of the HIN rate. While at \highmdot the peak luminosity is almost insensitive to the adopted carbon fusion rate, at \middmdot the THM rate leads to a luminosity peak of $\sim 80\%$ the value obtained with the CF88 rate.
% - - - - - - - - - - - - - - - - - - - - - - - - - - - - - - - - - - - - - - - - - - - - - - - - - - - - - %

% - - - - - - - - - - - - - - - - - - - - - - - - - - - - - - - - - - - - - - - - - - - - - - - - - - - - - %
To get more information as to how the carbon burning unfolds it is instructive to analyze the evolution of the temperature and the mass fractions of several isotopes along the path between \carbos{} and \iroh. Regarding the temperature, Fig.~\ref{fig:maximum_temperatures_1} displays the evolution of the maximum temperature for each of the four simulations as a function of time for both mass accretion rates, again shifting the time coordinate in each simulation to ensure that the luminosity peak of the explosion occurs at $t = 0$. Regardless of the accretion rate we observe both CF88 and HIN-RES simulations following an almost indistinguishable track.
In both panels of Fig.~\ref{fig:maximum_temperatures_1}, we observe a similar trend: after reaching a global maximum, there are two stages of temperature decrease. In the first, the temperature decrease is small, e.g. up to $\la 20\%$ of the peak value, and its duration is almost identical for the four simulations, $\sim 2$ hrs. In the second stage, the temperature drops by almost one order of magnitude, effectively returning to its pre-burst value, $10^{8.85}$~K at \middmdot\ and $\approx 10^{8.9}$~K at \highmdot. 
At \middmdot\ we observe that both CF88 and HIN-RES simulations reach an absolute maximum of $10^{9.6} \approx 4 \times 10^{9}$~K and a timescale of a few hours for the first drop in temperature. On the other hand, the THM-simulation exhibits a reduced absolute maximum of $\approx 3\times 10^{9}$~K and the fastest drop on the second stage of temperature decay of $\sim 10^{2}$ hr. In contrast, the HIN rate leads to  the largest peak temperature among the four models, $\sim 5\times 10^{9}$ K, as well as the slowest temperature decrease.
Let us remark that although the pre-bursting temperatures are rather similar for all four reaction rates at both \middmdot\ and \highmdot, the maximum column depth the fuel reaches before the explosion is different. As reported in Table~\ref{tab:table_1_superbursts_parameters_a56}, at both mass accretion rates the ignition is triggered at the shallowest column depth in the THM simulation and at the deepest column depth in the HIN simulation, while CF88 and HIN-RES yield intermediate values. In particular, these two cases exhibit similar ignition column depths, $y_{\mathrm{ignition}}\approx 4\times 10^{11}$ g cm$^{-2}$ at \middmdot\ and $\approx 10^{11}$ g cm$^{-2}$ at \highmdot.
This is explicitly visible as well in Fig.~\ref{fig:maximum_temperatures_2}. Here we show as contours the temperature of the envelope and crust as a function of column depth and time for the HIN-RES simulation. The dashed line, indicating the instantaneous column depth $y_c$ at which the mass fraction of \carbos{} is equal to $10^{-6}$, serves as a tracer to indicate the separation between accreted matter, at $y < y_{c}$, and the mixture between the ashes from the superburst and the previous composition at $y > y_{c}$. In both cases, we observe the maximum temperature is restricted to a range of $\log_{10}y\sim 11.5$ close to the explosion site. From this plot, we can see why the global maximum in temperature displays a two-stage drop: during the first of these, lasting $\sim 1$ hr, nuclear heating is sufficient as to counterbalance gravity. The high temperature is slowly propagated from the explosion site to the surface of the envelope, allowing incoming material to be burnt at the same rate at which it is accreted.

To better understand how a decreased peak temperature changes the nucleosynthesis at \middmdot, we show in Fig.~\ref{fig:mass_fractions_1} the time evolution of the average mass fraction of some representative nuclides obtained with the THM and HIN-RES rates (for the calculation of the averages, see Appendix~\ref{appendix:mass_fraction_computation}). In these figures, time has been shifted so that the luminosity peak of the superburst occurs at $t=0$.
In the upper left panel, we observe that \helioos, $^{16}$O and $^{24}$Mg are produced in similar amounts for both models.
The subsequent burning rate, however, is different: while \helioos{} undergoes a rapid decrease in the THM simulation, the $^{16}$O burning is actually inefficient as its average mass fraction quickly reaches an equilibrium point. On the other hand, $^{16}$O is burnt in large amounts in the HIN-RES simulation, reaching an equilibrium value about 100 times smaller than in the THM simulation.

If we look at the maximum temperature of the envelope in Fig.~\ref{fig:maximum_temperatures_1}, we notice that the THM simulation reaches a peak temperature of  $\sim3.5$ GK, whereas with the HIN-RES rate a maximum of
$\sim4.5$ GK can be reached. Such a difference is sufficient to modify the nuclear quasi-equilibrium established among the $\alpha$-chain nuclei. In the THM simulation, charged-particle reactions become inefficient
earlier during the cooling phase, leading to an earlier freeze-out and thus to larger residual abundances of intermediate $\alpha$-nuclides. For instance, the mass fraction of $^{16}$O freezes out at $\sim10^{-3}$ in the THM simulation, while it decreases to $\sim10^{-5}$ in the HIN-RES simulation.

In the upper right panel, showing $\alpha$-nuclides between $A = 28$ and $A=40$, we again observe similar amounts of them being produced during the temperature rise though the subsequent equilibrium mass fraction in the HIN-RES simulation is $\sim 10$ times smaller  than in the THM simulation. This trend is also visible for the mass fraction of $^{44}$Ti in the lower left panel of this same figure. However, for $^{48}$Cr and $^{50}$Cr we see a very different behavior since their decreasing abundances in both simulations are very similar. Such trend is also visible in the lower right panel for the mass fractions of several iron isotopes and for the last $\alpha$-nuclide in the network, $^{56}$Ni. Considering the difference in temperature is between 0.1 and 0.2 in $\log_{10}T$, as visible in Figs.~\ref{fig:maximum_temperatures_1} and~\ref{fig:maximum_temperatures_2}, we can thus attribute this similarity to the production of these heavy nuclides via $^{56}$Fe photodisintegration and $\alpha$-captures.
% - - - - - - - - - - - - - - - - - - - - - - - - - - - - - - - - - - - - - - - - - - - - - - - - - - - - - %

% - - - - - - - - - - - - - - - - - - - - - - - - - - - - - - - - - - - - - - - - - - - - - - - - - - - - - %
Fig.~\ref{fig:distribution_ashess_1} displays the average molar fraction $\langle Y \rangle = \langle X/A \rangle$ of the ashes as a function of mass $A$ and charge $Z$ numbers for all four carbon fusion rates and for both mass accretion rates. Due to the expensive computationally cost of the simulations at \middmdot, these ashes were computed after the end of the first superburst in the case of \middmdot and after the end of the second superburst for the \highmdot case. Here, we compute the mass average fractions $\langle X \rangle $ for each individual nuclide in the network and add them up according to their mass or charge number. Further details on the computation of these averages can be found in Appendix~\ref{appendix:mass_fraction_computation}.

The first characteristic to note from both panels is the invariance of the double-peaked structure with respect to variations in the mass accretion rate. We also observe the emergence of secondary maxima at $\alpha$-nuclides, such as $^{28}$Si or $^{36}$Ar. These, we can argue, are consequences of burning the same chemical composition albeit in slightly different conditions of temperature ($\sim 10^{9 - 9.5}$) and depth ($y\sim 10^{10-11}$ g cm$^{-2}$). Such features restrict the nucleosynthesis path that nuclides can follow. For instance, the presence of large amounts of isotopes with $56>A\geq 52$ in amounts that are almost insensitive to changes in both the carbon fusion rate and the mass accretion rate can be accounted by a combination of two paths. The first is the photodisintegration of $^{56}$Fe into isotopes such as $^{52}$Cr or $^{55}$Mn, which can actually proceed due to the large temperatures in excess of $10^{9}$~K reached during the superburst and the accreted material being $80\%$ composed of $^{56}$Fe. The second path is the $p$- and $\alpha$-captures by nuclides with $28 \leq A \leq 44$, once sufficient amounts of these nuclei have been produced from carbon fusion.

At \middmdot, we observe a substantial sensitivity on the distribution of ashes to changes in the reaction rate for $A\leq 44$ nuclides. In this region, we observe that the THM-boosted rate at $10^{9}$~K leads to a significantly larger amount of $\alpha$-nuclides with respect to the CF88/HIN-RES simulations, while the low HIN rate gives rise to lower abundances.

A striking characteristic of the four models at \highmdot is the presence of a substantial amount of nuclei with $28\leq A\leq 50$ in abundances larger than in the \middmdot scenario by almost a factor of $10$. In particular, we observe at \highmdot a large amount of $^{16}$O: in the right panels of Figs.~\ref{fig:distribution_ashess_1} we see average molar fraction of $\langle Y \rangle\sim 10^{-4}$, hence the average mass fraction corresponds to $\langle X \rangle = 16\times 10^{-4}$, i.e. $\sim 0.1\%$ of the total composition. The enhanced presence of $\alpha$-nuclides in the interval $28\leq A\leq 50$ at \highmdot can be explained as a combination of two factors. First of all, the evolution of the maximum temperature: if we look at both panels of Fig.~\ref{fig:maximum_temperatures_1}, we can notice all four simulations at \highmdot exhibit profiles that are similar to the one for the THM simulation at \middmdot regarding the absolute maximum temperature and the decay phases. This, in principle, helps explaining why we observe rather similar distribution of ashes when comparing the THM simulation in the left panel of Fig.~\ref{fig:distribution_ashess_1} with the ashes from the four simulations in the right panel of the same figure.
The second factor accounting for this enhancement of $28\leq A\leq 48$ nuclei is the high mass accretion rate, which leads to higher temperatures being reached at shallower column depths, as we can see in Table~\ref{tab:table_1_superbursts_parameters_a56} where $y_{\mathrm{ignition}}$ is comparable among the THM simulation at \middmdot with the four simulations at \highmdot.
Therefore, while physical conditions allow for a quick build-up of $A\leq 44$ isotopes, the fast decay of the peak temperature inhibits the nucleosynthesis of $A>44$ isotopes due to sole carbon burning.

%%--%%--%%--%%--%%--%%--%%--%%--%%--%%--%%--%%--%%--%%--%%--%%--%%--%%--%%--%%--%%--%%--%%--%%--%%--%%--%%--%%--%%--%%--%%--%%--%%--%%--%%--%%--%%--%%--%%--%%--%%--%%--%%--%%--%%--%%--%%--%%--%%--%%--%%--%%--%%--%%--%%--%%--%%
%%--%%--%%--%%--%%--%%--%%--%%--%%--%%--%%--%%--%%--%%--%%--%%--%%--%%--%%--%%--%%--%%--%%--%%--%%--%%--%%--%%--%%--%%--%%--%%--%%--%%--%%--%%--%%--%%--%%--%%--%%--%%--%%--%%--%%--%%--%%--%%--%%--%%--%%--%%--%%--%%--%%--%%--%%

%%--%%--%%--%%--%%--%%--%%--%%--%%--%%--%%--%%--%%--%%--%%--%%--%%--%%--%%--%%--%%--%%--%%--%%--%%--%%--%%--%%--%%--%%--%%--%%--%%--%%--%%--%%--%%--%%--%%--%%--%%--%%--%%--%%--%%--%%--%%--%%--%%--%%--%%--%%--%%--%%--%%--%%--%%
%%--%%--%%--%%--%%--%%--%%--%%--%%--%%--%%--%%--%%--%%--%%--%%--%%--%%--%%--%%--%%--%%--%%--%%--%%--%%--%%--%%--%%--%%--%%--%%--%%--%%--%%--%%--%%--%%--%%--%%--%%--%%--%%--%%--%%--%%--%%--%%--%%--%%--%%--%%--%%--%%--%%--%%--%%
\subsection{Base heating vs reaction rate version}\label{subsec:qb_vs_rate}
%%--%%--%%--%%--%%--%%--%%--%%--%%--%%--%%--%%--%%--%%--%%--%%--%%--%%--%%--%%--%%--%%--%%--%%--%%--%%--%%--%%--%%--%%--%%--%%--%%--%%--%%--%%--%%--%%--%%--%%--%%--%%--%%--%%--%%--%%--%%--%%--%%--%%--%%--%%--%%--%%--%%--%%--%%

%----------------------------------------------------------------------------------------------------------------------------------%
\begin{figure}[ht!]
    \centering
    \includegraphics[width=0.4\textwidth]{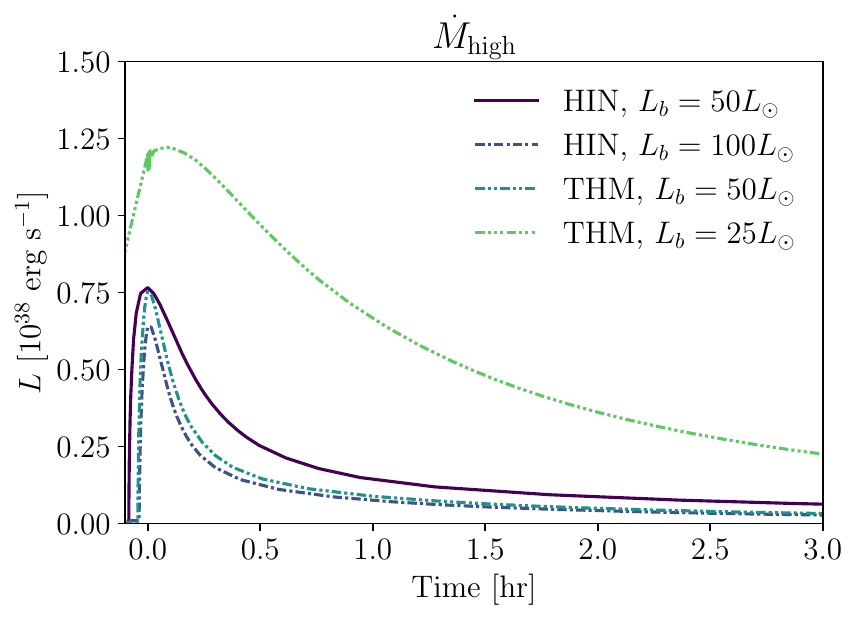}
    \caption{Same caption as in Fig.~\ref{fig:luminosity_profiles_1}, now for simulations at \highmdot with HIN and THM rates but at different base luminosity.}
    \label{fig:luminosity_curve_2}
\end{figure}
%----------------------------------------------------------------------------------------------------------------------------------%

To assess whether changing the reaction rate has a similar effect as changing the base luminosity, we ran two more simulations at \highmdot{}. We considered different values of $L_b$ and the THM and HIN versions of the carbon fusion rate, as they exhibit the largest discrepancies with respect to the reference CF88. Since the THM rate has a boosted value at $T < 10^{9}$~K, we reduced by half the base luminosity. For the HIN rate, having a decreased value at the same temperature interval, we doubled the value of $L_b$. In terms of $Q_b$, it means going from 0.16 to 0.32 MeV per baryon in the case of the HIN-simulation and from 0.16 to 0.08 MeV per baryon in the case of the THM-simulation. In Fig.~\ref{fig:luminosity_curve_2}, we display the luminosity profile of one superburst for each of these four simulations. Here we observe that doubling $L_b$ for a HIN-simulation leads to a comparable effect as to simply boosting the reaction rate but at the same $L_b$. Similarly, cutting by half the base luminosity in a THM-simulation has a comparable effect as to just reduce the carbon fusion rate. 
Regarding the recurrence time, the HIN and THM simulations at $L_b = 50L_{\odot}$ yield $t_{\mathrm{recurrence}}\sim 11$ and $\sim 21$ days respectively. 
For the HIN simulation at twice this $L_b$ value we find a recurrence time of 9.3 days. In contrast, for the THM simulation at half this same $L_b$ we find $t_{\mathrm{recurrence}} = 48.2$ days.

%%--%%--%%--%%--%%--%%--%%--%%--%%--%%--%%--%%--%%--%%--%%--%%--%%--%%--%%--%%--%%--%%--%%--%%--%%--%%--%%--%%--%%--%%--%%--%%--%%--%%--%%--%%--%%--%%--%%--%%--%%--%%--%%--%%--%%--%%--%%--%%--%%--%%--%%--%%--%%--%%--%%--%%--%%
%%--%%--%%--%%--%%--%%--%%--%%--%%--%%--%%--%%--%%--%%--%%--%%--%%--%%--%%--%%--%%--%%--%%--%%--%%--%%--%%--%%--%%--%%--%%--%%--%%--%%--%%--%%--%%--%%--%%--%%--%%--%%--%%--%%--%%--%%--%%--%%--%%--%%--%%--%%--%%--%%--%%--%%--%%

%%--%%--%%--%%--%%--%%--%%--%%--%%--%%--%%--%%--%%--%%--%%--%%--%%--%%--%%--%%--%%--%%--%%--%%--%%--%%--%%--%%--%%--%%--%%--%%--%%--%%--%%--%%--%%--%%--%%--%%--%%--%%--%%--%%--%%--%%--%%--%%--%%--%%--%%--%%--%%--%%--%%--%%--%%
%%--%%--%%--%%--%%--%%--%%--%%--%%--%%--%%--%%--%%--%%--%%--%%--%%--%%--%%--%%--%%--%%--%%--%%--%%--%%--%%--%%--%%--%%--%%--%%--%%--%%--%%--%%--%%--%%--%%--%%--%%--%%--%%--%%--%%--%%--%%--%%--%%--%%--%%--%%--%%--%%--%%--%%--%%
\subsection{On the impact of choosing $^{56}$Fe as a fiducial inert nuclide}\label{subsec:fe56_ge72}
%%--%%--%%--%%--%%--%%--%%--%%--%%--%%--%%--%%--%%--%%--%%--%%--%%--%%--%%--%%--%%--%%--%%--%%--%%--%%--%%--%%--%%--%%--%%--%%--%%--%%--%%--%%--%%--%%--%%--%%--%%--%%--%%--%%--%%--%%--%%--%%--%%--%%--%%--%%--%%--%%--%%--%%--%%

%----------------------------------------------------------------------------------------------------------------------------------%
\begin{figure}[ht!]
    \centering
    \includegraphics[width=0.47\textwidth]{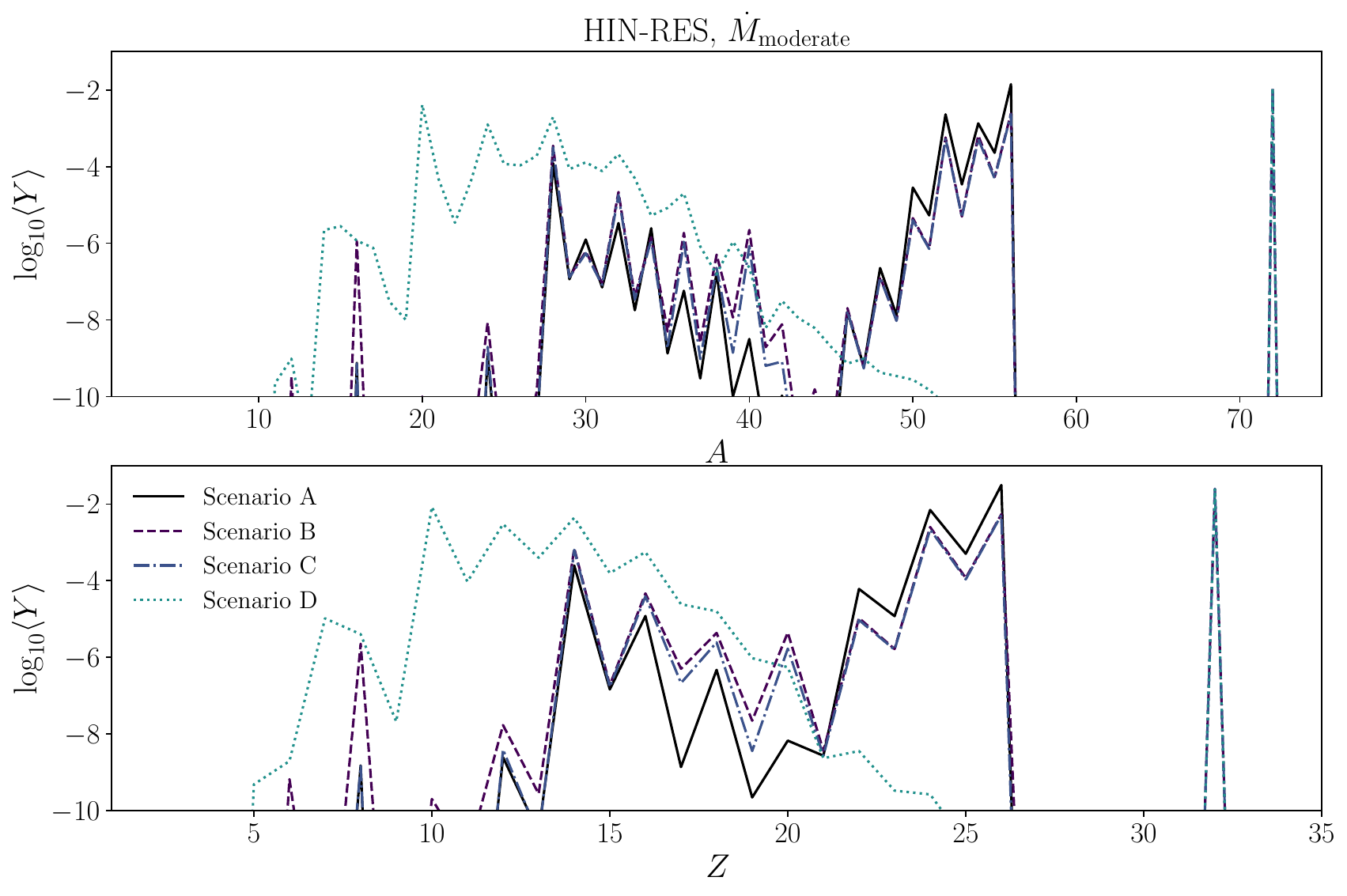}
    \includegraphics[width=0.47\textwidth]{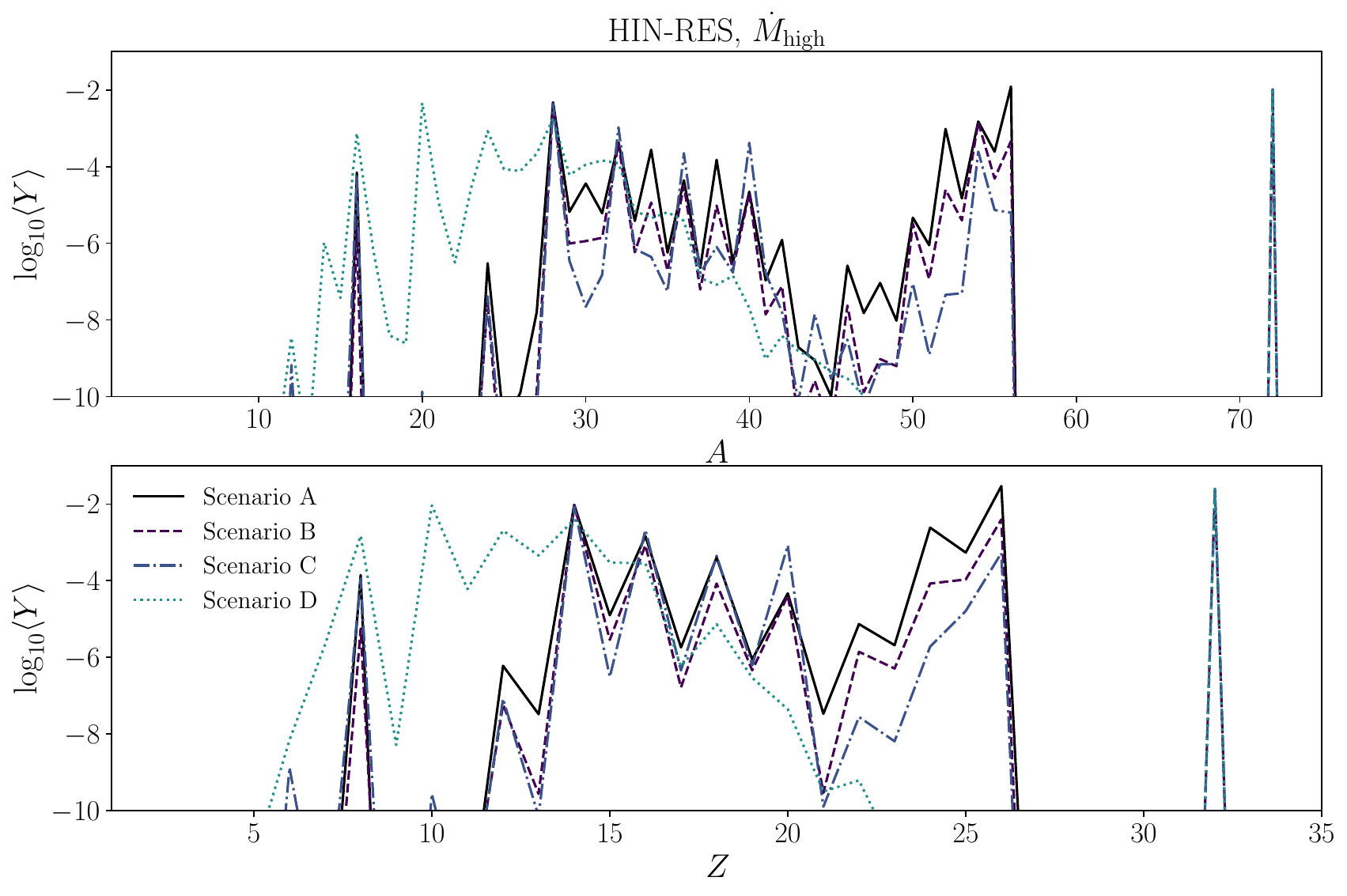}
    \caption{Same as in Fig.~\ref{fig:distribution_ashess_1}, for the four scenarios with the HIN-RES rate in which we vary the chemical composition of the accreted high-$Z$ nuclide and of the original envelope. Upper panels: simulations at \middmdot. Lower panels: simulations at \highmdot. See main text for further details.}
    \label{fig:distribution_ashess_1b}
\end{figure}
%----------------------------------------------------------------------------------------------------------------------------------%

%----------------------------------------------------------------------------------------------------------------------------------%
\begin{figure*}[ht!]
    \sidecaption
    \includegraphics[width=0.7\textwidth]{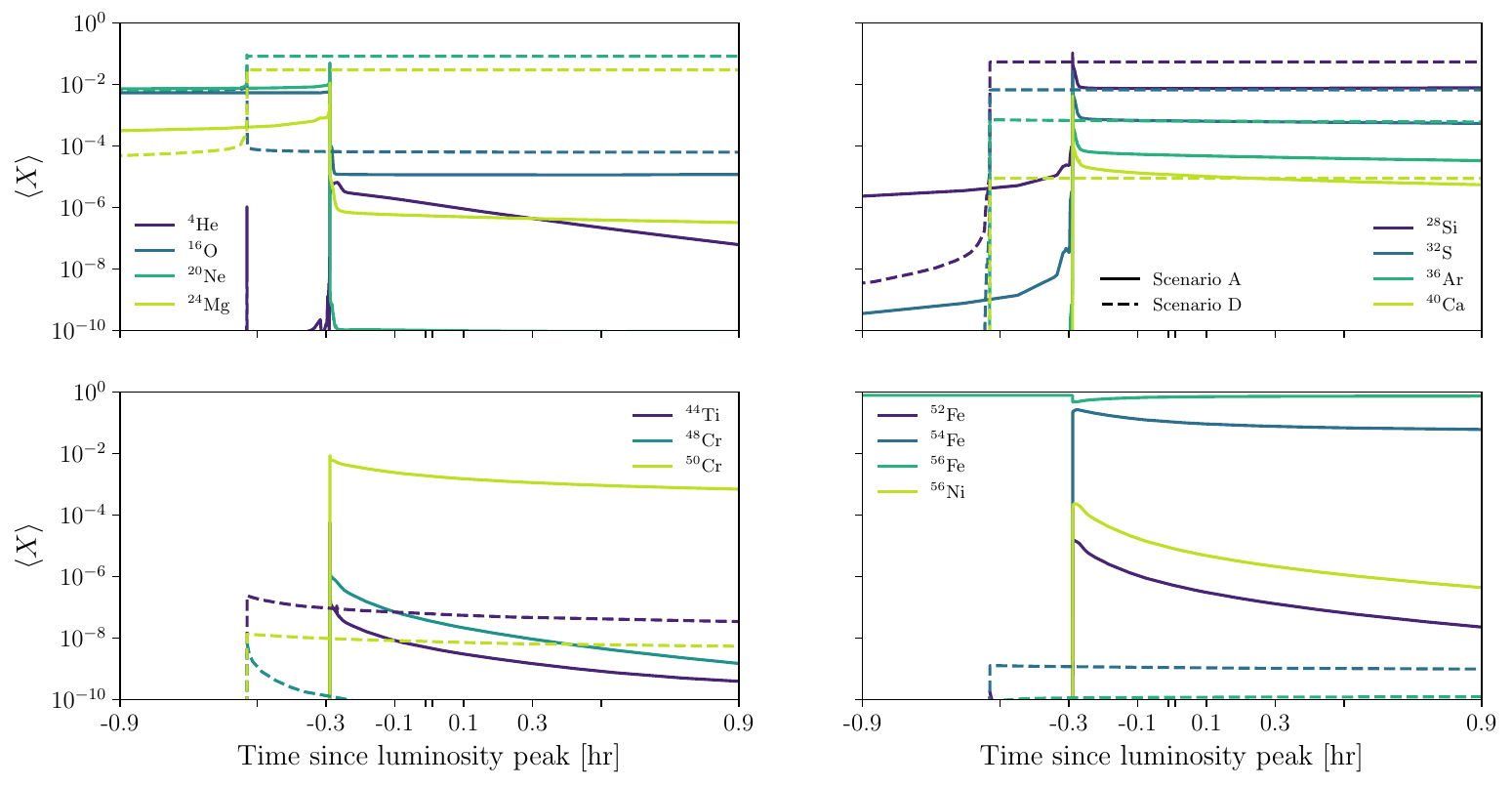}
    \caption{Same as Fig.~\ref{fig:mass_fractions_1}, now comparing the simulation with HIN-RES rate with the simulation where all photodisintegration reactions are removed (Scenario D).}
    \label{fig:mass_fractions_2}
\end{figure*}
%----------------------------------------------------------------------------------------------------------------------------------%

In Fig.~\ref{fig:distribution_ashess_1}, we observe the distribution of ashes reaching a peak at \iroh. It might be argued that such a peak is expected given the chemical composition of the original envelope and of the accreted material. Considering $T\gg 10^{9.4}\ \mathrm{K}$ in the envelope during a superburst, e.g. Figs.~\ref{fig:maximum_temperatures_1} and~\ref{fig:maximum_temperatures_2}, one might ponder what percentage of all $A\geq 50$ species abundances actually come from carbon burning instead of \iroh{} photodisintegration.
To address these questions, we explored three additional scenarios with the HIN-RES rate. We replaced \iroh{} in both the original composition of the envelope and in the amount of accreted material by a truly inert isotope from the point of view of the reaction network, i.e. not connected by any reaction to the rest of the species. For this role, we selected $^{72}$Ge, a choice based on the typical byproducts of rp-process ashes peaking at $A = 64$ and $A = 72$. The following scenarios were considered:
\begin{itemize}
    \item Scenario A: The HIN-RES simulations introduced in Sect. \ref{subsec:approx_a56}.
    \item Scenario B: The original envelope is composed of $^{56}$Fe while the accreted matter consists of $80\%$ $^{72}$Ge and $20\%$ \carbos;
    \item Scenario C: The original envelope is composed of pure $^{72}$Ge and the accreted material composition is divided into $80\%$ $^{72}$Ge and $20\%$ \carbos, 
    \item Scenario D: Same as Scenario C, but removed all photodisintegration reactions from the network.
\end{itemize}
Figure~\ref{fig:distribution_ashess_1b} shows the resulting distributions of ashes from these four scenarios after a superburst at both \middmdot and \highmdot.
Among Scenarios A, B and C, we observe a relative good agreement in the distribution of abundances of $^{16}$O, \silicos{} and several other $\alpha$-nuclides, regardless of the mass accretion rate. While the overall double-peaked distribution seems independent of whether we choose $^{56}$Fe or $^{72}$Ge as the inert high-$Z$ nuclide in the accreted material, the end-point of the nucleosynthesis is different as a consequence of changing the mass accretion rate as it is $^{56}$Fe at \middmdot while at \highmdot the endpoint is at $^{54}$Fe.

At \middmdot, the distribution of ashes around the iron peak in Scenarios A, B and C is rather similar, while at \highmdot we observe an underproduction of $A = 50$ and $A = 52$ nuclei if $^{56}$Fe is absent in the accreted material. However, the distribution of ashes between $A = 28$ and $A = 40$ is overly similar among Scenarios A, B and C regardless of the mass accretion rate, which is consistent with the facts these nuclides are mainly synthesized by \carbos{} burning and the same amount of \carbos{} is accreted in these three scenarios.

The major differences in the distribution of ashes among Scenarios A, B, and C occur in the intermediate-mass region, around $A\sim 40$ (from S to Ca) and corresponding approximately to nuclei with $16 \leq Z \leq 26$.
At \middmdot, Scenario A shows lower abundances for nuclei with $Z\leq 20$ and enhanced production for $Z>20$, whereas the opposite trend is found in Scenarios B and C. At \highmdot, on the other hand, we obtain good agreement among the three scenarios for $Z\leq 20$, while for $Z>20$ the production of heavy nuclei appears to be reduced in Scenarios B and C, possibly as a consequence of the different seed composition of the accreted material (e.g. $^{56}$Fe in Scenario A versus $^{64}$Zn in Scenarios B and C).

Scenario D, on the other hand, exhibits a very different distribution of ashes in comparison to those from Scenarios A, B and C. While at both \middmdot and \highmdot we still observe similar amounts of $^{16}$O and \silicos{} in all four Scenarios - a consequence of $^{12}$C($^{12}$C, $\alpha$)$^{20}$Ne and the subsequent $^{12}$C($\alpha,\gamma$)$^{16}$O and $^{16}$O($^{16}$O,$\alpha$)$^{28}$Si -  we observe that by removing all photo-disintegration channels from the network the nucleosynthesis cannot proceed beyond $^{40}$Ca. Instead, we see large concentrations of $^{20}$Ne and $^{24}$Mg, both byproducts of \carbos{} burning, comparable with the concentration of \silicos. Such an enhancement is independent of the mass accretion rate selected. The minor peaks of \sulfuros{} and $^{36}$Ar are consistent with what we previously observed in the nucleosynthesis during the explosion, e.g. Fig.~\ref{fig:mass_fractions_1}, and helps clarifying the nucleosynthesis for Scenario D: while $p-$ and $\alpha-$capture reactions, such as $^{36}$Ar($\alpha,\gamma$)$^{40}$Ca, are favored in these temperature and density conditions, all the light elements generated during the early stages of carbon burning quickly exhaust the available supply of $p$ and $\alpha$ particles. The newly formed products of such captures, e.g. \neos{} and $^{24}$Mg, cannot release $p$ or $\alpha$ particles if photoreactions are turned off. This impedes the synthesis of $A\geq 44$ nuclides. If such photodisintegrations were present, as in Scenario C, $^{40}$Ca and $^{44}$Ti can be produced and, due to the high temperatures at the envelope (Fig.~\ref{fig:maximum_temperatures_2}), synthesis of $A\sim 54$ becomes plausible. This description of the nucleosynthesis can be explicitly seen in Fig.~\ref{fig:mass_fractions_2}, where we compare the average mass fraction for several nuclides of Scenarios A and D at \middmdot.

%%--%%--%%--%%--%%--%%--%%--%%--%%--%%--%%--%%--%%--%%--%%--%%--%%--%%--%%--%%--%%--%%--%%--%%--%%--%%--%%--%%--%%--%%--%%--%%--%%--%%--%%--%%--%%--%%--%%--%%--%%--%%--%%--%%--%%--%%--%%--%%--%%--%%--%%--%%--%%--%%--%%--%%--%%
%%--%%--%%--%%--%%--%%--%%--%%--%%--%%--%%--%%--%%--%%--%%--%%--%%--%%--%%--%%--%%--%%--%%--%%--%%--%%--%%--%%--%%--%%--%%--%%--%%--%%--%%--%%--%%--%%--%%--%%--%%--%%--%%--%%--%%--%%--%%--%%--%%--%%--%%--%%--%%--%%--%%--%%--%%

%%--%%--%%--%%--%%--%%--%%--%%--%%--%%--%%--%%--%%--%%--%%--%%--%%--%%--%%--%%--%%--%%--%%--%%--%%--%%--%%--%%--%%--%%--%%--%%--%%--%%--%%--%%--%%--%%--%%--%%--%%--%%--%%--%%--%%--%%--%%--%%--%%--%%--%%--%%--%%--%%--%%--%%--%%
%%--%%--%%--%%--%%--%%--%%--%%--%%--%%--%%--%%--%%--%%--%%--%%--%%--%%--%%--%%--%%--%%--%%--%%--%%--%%--%%--%%--%%--%%--%%--%%--%%--%%--%%--%%--%%--%%--%%--%%--%%--%%--%%--%%--%%--%%--%%--%%--%%--%%--%%--%%--%%--%%--%%--%%--%%
\subsection{Simulations with \approxaii\ at \highmdot}\label{subsec:approx_a64}
%%--%%--%%--%%--%%--%%--%%--%%--%%--%%--%%--%%--%%--%%--%%--%%--%%--%%--%%--%%--%%--%%--%%--%%--%%--%%--%%--%%--%%--%%--%%--%%--%%--%%--%%--%%--%%--%%--%%--%%--%%--%%--%%--%%--%%--%%--%%--%%--%%--%%--%%--%%--%%--%%--%%--%%--%%
%----------------------------------------------------------------------------------------------------------------------------------%
\begin{figure}[ht!]
    \centering
    \includegraphics[width=0.45\textwidth]{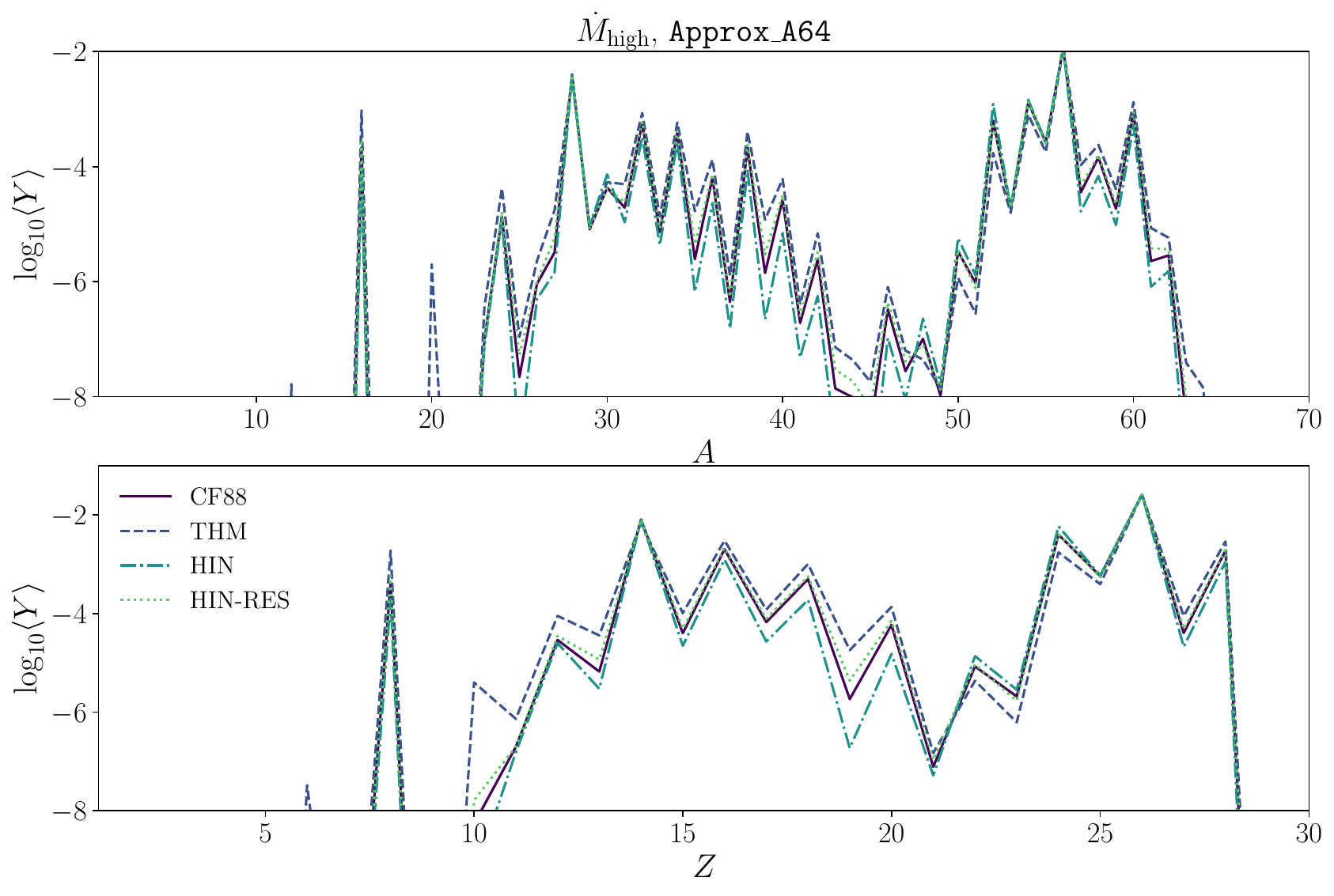}    
    \caption{Average abundances for the ashes of the superbursts, depicted as a function of mass number A (upper panel) and as a function of charge number Z (lower panel).}
    \label{fig:approxa64_ashes}
\end{figure}
%----------------------------------------------------------------------------------------------------------------------------------%

The distribution of ashes of a superburst at \middmdot is found to be in relative good agreement with the \kepler{} simulations. This suggests using an approximate network of $A_{\max} = 56$ is appropriate to simulate superbursts at \middmdot. However, at \highmdot we obtained large amounts of accumulated material between $A = 16$ and $A = 52$, which call out for examining the role of isotopes with $A>56$ on the properties of the superburst at this high mass accretion rate.
This was done by running four simulations at \highmdot with the extended \approxaii\ network. The resulting distributions of ashes as a function of both mass and charge number are shown in Fig.~\ref{fig:approxa64_ashes}: here we observe that even at such a high accretion rate as \highmdot the synthesis of nuclides is unlikely to proceed beyond $A = 64$. Still we do obtain two small abundance peaks below $10^{-6}$ at $A = 60$ and $A = 62$, indicating that the truncation at $A = 56$ is not strictly valid at such a high accretion rate. Since the accumulated material at $A = 64$ has an abundance of $\leq 10^{-8}$, the likelihood of producing isotopes with $A > 64$ remains small. In agreement with the \approxai{} simulations in Fig.~\ref{fig:distribution_ashess_1}, we still get about $0.1\%$ of $^{16}$O, indicating this is not an artifact of truncating the network at $A = 56$.
%%--%%--%%--%%--%%--%%--%%--%%--%%--%%--%%--%%--%%--%%--%%--%%--%%--%%--%%--%%--%%--%%--%%--%%--%%--%%--%%--%%--%%--%%--%%--%%--%%--%%--%%--%%--%%--%%--%%--%%--%%--%%--%%--%%--%%--%%--%%--%%--%%--%%--%%--%%--%%--%%--%%--%%--%%
%%--%%--%%--%%--%%--%%--%%--%%--%%--%%--%%--%%--%%--%%--%%--%%--%%--%%--%%--%%--%%--%%--%%--%%--%%--%%--%%--%%--%%--%%--%%--%%--%%--%%--%%--%%--%%--%%--%%--%%--%%--%%--%%--%%--%%--%%--%%--%%--%%--%%--%%--%%--%%--%%--%%--%%--%%

%%%%%%%%%%%%%%%%%%%%%%%%%%%%%%%%%%%%%%%%%%%%%%%%%%%%%%%%%%%%%%%%%%%%%%%%%%%%%%%%%%%%%%%%%%%%%%%%%%%%%%%%%%%%%%%%%%%%%%%%%%%%%%%%%%%%%%%%%%%%%%%%%%%%%%%%%%%%%%%%%%%%%%%%%%%%%%%%%%%%%%%%%%%%%%%%%%%%%%%%%%%%%%%%%%%%%%%%%%%%%%%%%%%
%%%%%%%%%%%%%%%%%%%%%%%%%%%%%%%%%%%%%%%%%%%%%%%%%%%%%%%%%%%%%%%%%%%%%%%%%%%%%%%%%%%%%%%%%%%%%%%%%%%%%%%%%%%%%%%%%%%%%%%%%%%%%%%%%%%%%%%%%%%%%%%%%%%%%%%%%%%%%%%%%%%%%%%%%%%%%%%%%%%%%%%%%%%%%%%%%%%%%%%%%%%%%%%%%%%%%%%%%%%%%%%%%%%

%%%%%%%%%%%%%%%%%%%%%%%%%%%%%%%%%%%%%%%%%%%%%%%%%%%%%%%%%%%%%%%%%%%%%%%%%%%%%%%%%%%%%%%%%%%%%%%%%%%%%%%%%%%%%%%%%%%%%%%%%%%%%%%%%%%%%%%%%%%%%%%%%%%%%%%%%%%%%%%%%%%%%%%%%%%%%%%%%%%%%%%%%%%%%%%%%%%%%%%%%%%%%%%%%%%%%%%%%%%%%%%%%%%
%%%%%%%%%%%%%%%%%%%%%%%%%%%%%%%%%%%%%%%%%%%%%%%%%%%%%%%%%%%%%%%%%%%%%%%%%%%%%%%%%%%%%%%%%%%%%%%%%%%%%%%%%%%%%%%%%%%%%%%%%%%%%%%%%%%%%%%%%%%%%%%%%%%%%%%%%%%%%%%%%%%%%%%%%%%%%%%%%%%%%%%%%%%%%%%%%%%%%%%%%%%%%%%%%%%%%%%%%%%%%%%%%%%
\section{Discussion}\label{sec:sub_ending}
%%%%%%%%%%%%%%%%%%%%%%%%%%%%%%%%%%%%%%%%%%%%%%%%%%%%%%%%%%%%%%%%%%%%%%%%%%%%%%%%%%%%%%%%%%%%%%%%%%%%%%%%%%%%%%%%%%%%%%%%%%%%%%%%%%%%%%%%%%%%%%%%%%%%%%%%%%%%%%%%%%%%%%%%%%%%%%%%%%%%%%%%%%%%%%%%%%%%%%%%%%%%%%%%%%%%%%%%%%%%%%%%%%%

%%--%%--%%--%%--%%--%%--%%--%%--%%--%%--%%--%%--%%--%%--%%--%%--%%--%%--%%--%%--%%--%%--%%--%%--%%--%%--%%--%%--%%--%%--%%--%%--%%--%%--%%--%%--%%--%%--%%--%%--%%--%%--%%--%%--%%--%%--%%--%%--%%--%%--%%--%%--%%--%%--%%--%%--%%
%%--%%--%%--%%--%%--%%--%%--%%--%%--%%--%%--%%--%%--%%--%%--%%--%%--%%--%%--%%--%%--%%--%%--%%--%%--%%--%%--%%--%%--%%--%%--%%--%%--%%--%%--%%--%%--%%--%%--%%--%%--%%--%%--%%--%%--%%--%%--%%--%%--%%--%%--%%--%%--%%--%%--%%--%%
\subsection{On the general properties of superbursts and their ignition conditions}\label{subsec:discussion_1}
%%--%%--%%--%%--%%--%%--%%--%%--%%--%%--%%--%%--%%--%%--%%--%%--%%--%%--%%--%%--%%--%%--%%--%%--%%--%%--%%--%%--%%--%%--%%--%%--%%--%%--%%--%%--%%--%%--%%--%%--%%--%%--%%--%%--%%--%%--%%--%%--%%--%%--%%--%%--%%--%%--%%--%%--%%

%----------------------------------------------------------------------------------------------------------------------------------%
\begin{figure}[ht!]
    \centering
    \includegraphics[width=0.45\textwidth]{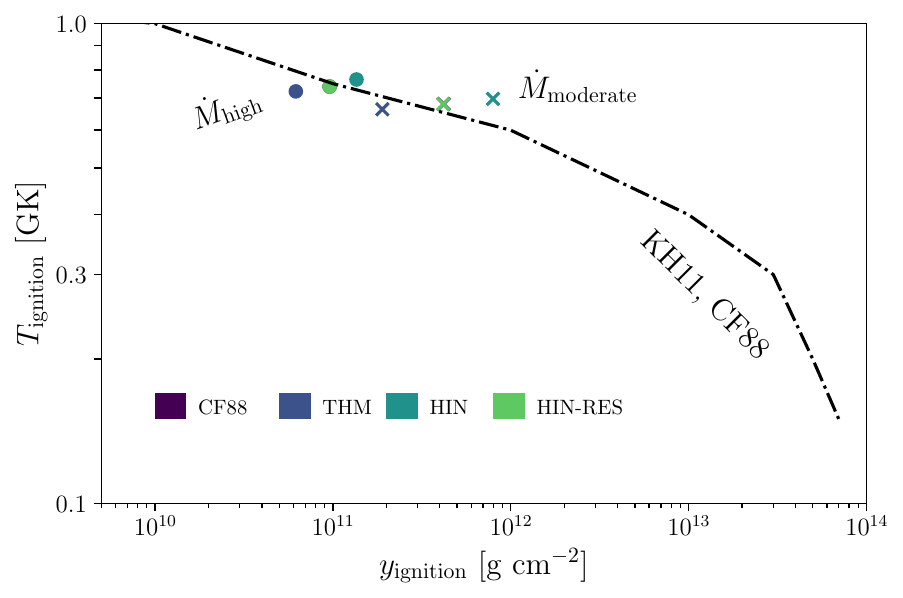}    
    \caption{Temperature and column depth at ignition for the superbursts in this work. The dash-dot line corresponds to the ignition curve shown in Fig.3 of KH11 for the CF88 reaction rate.}
    \label{fig:ignition_conditions}
\end{figure}
%----------------------------------------------------------------------------------------------------------------------------------%

% - - - - - - - - - - - - - - - - - - - - - - - - - - - - - - - - - - - - - - - - - - - - - - - - - - - - - %
Let us first discuss the overall characteristics of the superbursts and compare them with those from the \kepler\  multi-zone simulations of superbursts in KH11.
All of our simulations in Sect.~\ref{sec:results} were carried out at fixed values of $Q_b$, with appropriate values for each \middmdot and \highmdot in order to guarantee an unstable C-burning. As reported in Figs. 4 and 12 from \cite{KH2011} changing $Q_b$ and $\dot{M}$ has a similar effect of increasing/decreasing the timescales of the superburst.
Concerning the simulations at \middmdot, KH11 report that increasing the base heating from 0.17 to 0.27 MeV per baryon induces a decrease in the recurrence time of the superbursts from $\sim 328$ to $\sim 146$ days. Working at $Q_b = 0.26$ MeV per baryon in our models we find a similar window of changes: on the one hand, the THM-simulation displays a recurrence time of $\sim 100$ days shorter than the CF88 simulation. On the other hand, the recurrence time for the HIN-simulation is $\sim 364$ days, almost three times the value of the CF88 simulation. 
At \highmdot, a modification of $Q_b$ between 0.1 and 0.2 MeV per baryon induces less pronounced changes in the recurrence time, typically about $\sim 10^{-2}$ yr $\approx 3.6$ days. In contrast, in our simulations at fixed $Q_b$ we observe that at this same mass accretion rate the changes in the recurrence time are in the order of $\pm 5$ days between CF88 and the two extreme THM and HIN cases (see Table~\ref{tab:table_1_superbursts_parameters_a56}). A less pronounced effect is observed in the decay time of the luminosity, where modifying the reaction rate induces changes in the order of $\sim 0.2$ hr, comparable to the change induced by $Q_b$ of $\sim 0.1$ hr as reported by KH11. 

From these considerations, we observe that as long as the mass accretion rate of a system is well constrained, the timescales of the superbursts are equally sensitive to changes in the reaction rate at $T\leq 10^{9}$~K and to changes in the base luminosity $Q_b$. From an observational point of view, however, constraining the mass accretion rate is subject to several systematic uncertainties. In the standard approach, $\dot{M}$ is inferred from the observed X-ray flux and therefore depends on the estimated distance to the source and on the assumed neutron-star mass and radius \citep[e.g.][]{Page_2022}. This approach, however, does not take into account that the observed flux may be anisotropic due to the geometry of the accretion disk and its inclination angle \citep{He_Keek_2016, Galloway_2021}. In such cases, separating the contributions from direct emission and disk reflection may be challenging.

% - - - - - - - - - - - - - - - - - - - - - - - - - - - - - - - - - - - - - - - - - - - - - - - - - - - - - %

% - - - - - - - - - - - - - - - - - - - - - - - - - - - - - - - - - - - - - - - - - - - - - - - - - - - - - %
In a recent work, \citet{Meisel_2024} discussed the possibility of constraining the amount of shallow crustal heating through the ignition conditions of superbursts.
These conditions were determined from the crossing point of two temperature profiles: 
(i) the first is obtained from the one-zone model criterion for explosions, namely $d\varepsilon_\mathrm{heat}/dT = d\varepsilon_\mathrm{cool}/dT$) (for calculating these curves, they considered several versions of the carbon fusion rate, among them the THM and HIN versions),  (ii) the second is computed with the multi-zone crust cooling code \texttt{dStar}.
However, their calculations do not explicitly follow the thermonuclear evolution during the superburst itself, including carbon burning and the associated nucleosynthesis. 
In the diagrams of \citet{Meisel_2024}, it is clear that the strength of the shallow heating $Q_\mathrm{sh}$ plays a more prominent role than the deep crustal heating and/or the recently-proposed crust-Urca cooling mechanism.
Therefore, for the purpose of a qualitative comparison we regard both quantities as parametrizations of the net heat flux entering the envelope.
What we observe is that the ignition conditions obtained in our present calculations (see, e.g., Fig.~\ref{fig:ignition_conditions}) would require $Q_\mathrm{sh}\sim 1$ MeV per baryon in the analysis of \citet{Meisel_2024} regardless of the adopted carbon fusion rate.
On the other hand, their models at $Q_\mathrm{sh}\sim 0.1$ MeV per baryon comparable to our $Q_b$, demand an ignition depth $y_\mathrm{ignition}\sim 10^{13}-10^{14}$ g~cm$^{-2}$ that is about two orders of magnitude higher than the one we find. 
In contrast, all KH11 simulations at $\dot{M}$ comparable to our \middmdot{} and \highmdot require a $Q_b$ less than 0.3 MeV per baryon to obtain ignition conditions similar to ours. 
Moreover, the present and KH11 results suggest that invoking a strong shallow-heating source may not be necessary to reproduce the observed ignition properties of superbursts and that this conclusion may depend on the level of physical details included in the treatment of the thermonuclear burning.
Note that, by construction, both \mesa{} and \kepler{} take the heating from the deep layers of the crust as a mere input parameter in the form of a constant base luminosity $L_b$. Thus, neither \mesa{} nor \kepler{} are able to discern which processes in the crust contribute to the heating budget. At most, they are sensitive only to the net heat flux resulting from these processes. This, however, introduces an additional degeneracy since different physical processes in the crust may lead to similar effective values of this quantity.
In this regard, we are cautious about interpreting superburst simulations as evidence for specific heating or cooling mechanisms operating in the crust such as Urca cooling \citep{Meisel_2024, Huang_2026}.
In particular, our results show that changes in the adopted carbon fusion rate can produce effects on the superburst light curves comparable to those obtained by varying the base heating.

% - - - - - - - - - - - - - - - - - - - - - - - - - - - - - - - - - - - - - - - - - - - - - - - - - - - - - %

%%--%%--%%--%%--%%--%%--%%--%%--%%--%%--%%--%%--%%--%%--%%--%%--%%--%%--%%--%%--%%--%%--%%--%%--%%--%%--%%--%%--%%--%%--%%--%%--%%--%%--%%--%%--%%--%%--%%--%%--%%--%%--%%--%%--%%--%%--%%--%%--%%--%%--%%--%%--%%--%%--%%--%%--%%
%%--%%--%%--%%--%%--%%--%%--%%--%%--%%--%%--%%--%%--%%--%%--%%--%%--%%--%%--%%--%%--%%--%%--%%--%%--%%--%%--%%--%%--%%--%%--%%--%%--%%--%%--%%--%%--%%--%%--%%--%%--%%--%%--%%--%%--%%--%%--%%--%%--%%--%%--%%--%%--%%--%%--%%--%%

%%--%%--%%--%%--%%--%%--%%--%%--%%--%%--%%--%%--%%--%%--%%--%%--%%--%%--%%--%%--%%--%%--%%--%%--%%--%%--%%--%%--%%--%%--%%--%%--%%--%%--%%--%%--%%--%%--%%--%%--%%--%%--%%--%%--%%--%%--%%--%%--%%--%%--%%--%%--%%--%%--%%--%%--%%
%%--%%--%%--%%--%%--%%--%%--%%--%%--%%--%%--%%--%%--%%--%%--%%--%%--%%--%%--%%--%%--%%--%%--%%--%%--%%--%%--%%--%%--%%--%%--%%--%%--%%--%%--%%--%%--%%--%%--%%--%%--%%--%%--%%--%%--%%--%%--%%--%%--%%--%%--%%--%%--%%--%%--%%--%%
\subsection{On the distribution of ashes from superbursts}\label{subsec:discussion_2}
%%--%%--%%--%%--%%--%%--%%--%%--%%--%%--%%--%%--%%--%%--%%--%%--%%--%%--%%--%%--%%--%%--%%--%%--%%--%%--%%--%%--%%--%%--%%--%%--%%--%%--%%--%%--%%--%%--%%--%%--%%--%%--%%--%%--%%--%%--%%--%%--%%--%%--%%--%%--%%--%%--%%--%%--%%

%----------------------------------------------------------------------------------------------------------------------------------%
\begin{figure}[ht!]
    \centering
    \includegraphics[width=0.47\textwidth]{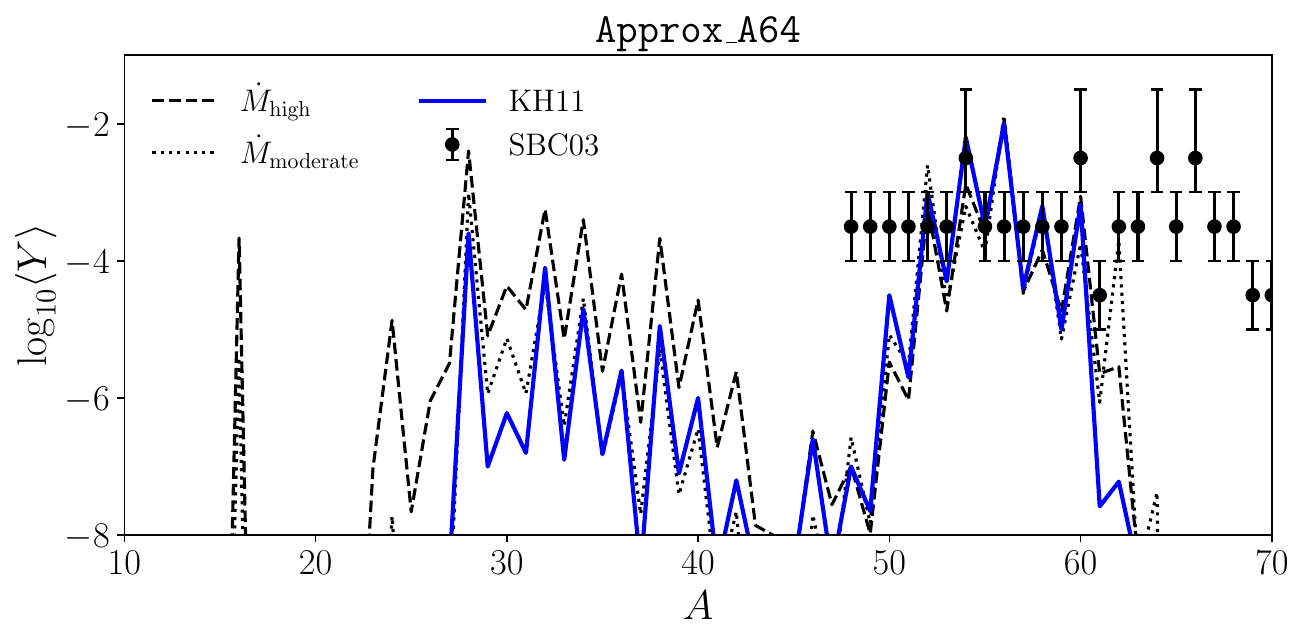}
    \caption{Same caption as in Fig.~\ref{fig:distribution_ashess_1}, now for CF88 models at \middmdot{} and \highmdot with the \approxaii{} network, calculated with \mesa{}, and the distributions of ashes obtained by KH11 (blue solid line) and SBC03 \cite[symbols;][]{Schatz_2003}. See text for further details.}
    \label{fig:distribution_ashess_compar}
\end{figure}
%----------------------------------------------------------------------------------------------------------------------------------%

% - - - - - - - - - - - - - - - - - - - - - - - - - - - - - - - - - - - - - - - - - - - - - - - - - - - - - %
Regarding the distribution of ashes for the eight models in Fig.~\ref{fig:distribution_ashess_1}, the first feature we observe is the qualitative agreement with the distribution of ashes reported in the literature \citep{Lau_2018, Jain_2025}. We are able to reproduce the doubled-peaked structure with local maxima at $A=28$ and $A = 56$ and the drop in abundances at $A\sim 40$, as well as the secondary maxima for $\alpha$ nuclides.

Aiming at a more in-depth discussion, in Fig. ~\ref{fig:distribution_ashess_compar} we explicitly compare the distribution of ashes of our \mesa{} simulations with two datasets available from the literature, namely KH11 and \citet{Schatz_2003}, hereafter SBC03. The first dataset corresponds to the distribution shown in the diagrams presented by \cite{Lau_2018, Jain_2025}, which are attributed to the superburst ashes calculated in KH11. This latter paper, however, does not explicitly provide the corresponding distribution of ashes in tabulated or graphical form. Nevertheless, these ashes have been adopted in subsequent works as representative superburst ashes \citep[e.g.][]{Lau_2018, Shchechilin_2021, Shchechilin_2022, Jain_2025}. The accretion rate and base luminosity of the underlying model are not specified in these works. Based on the available information, we infer that the employed \kepler{} calculations included nuclei with $A\leq 64$ and assumed an accreted composition of $20\%$ \carbos{} and $80\%$ \iroh{}.
The second dataset corresponds to the one-zone calculation presented by SBC03. The abundance distribution in SBC03 is reported graphically as colored squares indicating upper and lower abundance limits rather than as numerical values. Therefore, for the purpose of this comparison, we adopt the midpoint of the reported ranges as representative abundance values and use the corresponding intervals as uncertainties.
For our \mesa{} simulations, we show the ashes distributions obtained at \middmdot{} and \highmdot{} using the default CF88 reaction rate and the \approxaii{} network.
With respect to KH11, regardless of the selected mass accretion rate we observe three common characteristics. First, the existence of a double-peaked distribution of ashes with local maxima at $A = 28$ and $A=56$ and for various $\alpha$-nuclides, as well as a global minimum at $A = 44$. The second characteristic is the synthesis of nuclides with $A>56$  with the last prominent local maxima at $A = 60$. The third common featrure is the comparable values of the abundances.
At \highmdot{}, in contrast to the simulation at \middmdot{}, we find traces of isotopes below $A = 28$.
Regarding SBC03, we find an agreement in several peaks at $A\geq 50$. Notice their distribution of ashes suggest the existence of a third peak in the region $A = 60-70$, very likely a consequence of the photodisintegration of $^{68}$Zn and $^{104}$Ru and the subsequent $\alpha-$captures over the products of such photodisintegration. However, SBC03 did not find the first peak at $A = 28-40$, which can be a consequence of different factors, such as (i) starting with only $10\%$ of \carbos{}, (ii) considering no $A \sim 56$ material in the initial composition, despite having included some traces of $^{24}$Mg and \silicos, (iii) being a one-zone model. It would be of particular interest to extend the SBC03 calculation to a multi-zone simulation, i.e. to explore the actual consequences of burning an initial chemical composition based on the distribution of rp-process ashes within a multi-zone model. Such calculation could clarify whether the presence of the $A=28-40$ peak is an intrinsic outcome of carbon burning or instead depends sensitively on the assumed initial composition.
% - - - - - - - - - - - - - - - - - - - - - - - - - - - - - - - - - - - - - - - - - - - - - - - - - - - - - %

% - - - - - - - - - - - - - - - - - - - - - - - - - - - - - - - - - - - - - - - - - - - - - - - - - - - - - %
Most studies on superburst ashes take as basis point the distribution reported in KH11 and subsequent references. Considering the composition of the accreted material is artificially enriched in \iroh{}, one might ponder whether the distribution of isotopes near Fe is either a consequence of such an initial enrichment followed by $\alpha$- or $p$-captures or is in fact resulting from the carbon fusion. In Sect.~\ref{subsec:fe56_ge72} we have shown that, even by accreting a truly inert nuclide with respect to the selected network, the peaks at $A = 50, 52, 54$ and $56$ are still present albeit at proportions $\sim 10$ times smaller than what it is usually reported. In particular, we observe that $A = 54$ represents the actual endpoint of the process. 

A similar question was explored by SBC03 in the framework of a one-zone model. Using an initial composition representative of X-ray burst ashes, i.e. including nuclides such as $^{68}$Zn and $^{104}$Ru and setting the mass fraction of $^{12}$C to 0.1, they followed the thermal evolution of this material with a reaction network (neglecting electroweak reactions). It was shown that heavy nuclei from typically $^{68}$Zn up to $^{104}$Ru may photodisintegrate during the early stage of the superburst, thereby enriching the $A=56$ and $A=64$ isobaric groups. The final abundance distribution was obtained by assuming nuclear statistical equilibrium at fixed temperature and density.
In our multi-zone calculations, we obtain maximum temperatures between 2 and 6 GK, whereas the SBC03 one-zone calculation reaches temperatures close to 7 GK. The temperature adopted in their nuclear statistical equilibrium calculations, $T=4.2$ GK, lies within the range of temperatures reached in our models at \middmdot{} after the luminosity peak (e.g. Fig.~\ref{fig:maximum_temperatures_1}).
The resulting distributions of ashes, however, may differ quantitatively from both the present calculations and those inferred from KH11. Such differences likely reflect the sensitivity of the final abundances to the assumed initial composition and to the treatment of the cooling and freeze-out phases.
% - - - - - - - - - - - - - - - - - - - - - - - - - - - - - - - - - - - - - - - - - - - - - - - - - - - - - %

% - - - - - - - - - - - - - - - - - - - - - - - - - - - - - - - - - - - - - - - - - - - - - - - - - - - - - %
Most discussions of superbursts focus on sources with $\dot{M}\sim 0.3\dot{M}_\mathrm{Edd}$, close to our \middmdot. There is, however, an intriguing exception. Namely, the source GX 17+2 \citep{intZand_2004, intZand_2017}, for which all its four superbursts have been reported to occur at mass accretion rates close to $\dot{M}_{\mathrm{Edd}}$. The characteristics of these events are as follows: recurrence time of $\sim 10$ days, maximum luminosity between $1-2\ \times 10^{38}\, \mathrm{erg\ s}^{-1}$ and average decay time of $\sim 1.5$ hr. These values are, in a broad sense, consistent with our simulations at $2\times 10^{-8}\, M_{\odot}\, \mathrm{yr}^{-1}$ although the reported characteristics seem to favor a scenario for which the carbon fusion reaction rate is slightly less enhanced at low temperatures, as in the THM-simulation. 
In this regard, GX 17+2 provides an interesting observational counterpart for our simulations at high mass accretion rate. According to our simulations, a high accretion rate leads to an enhancement of $A\leq 40$ material, as well as to the synthesis of $\sim 0.1\%$ of $^{16}$O. While it might be argued that such an accumulation of low-$Z$ material is an artifact of the network - i.e. an incomplete burning of carbon which could push the nucleosynthesis up to $A\geq 64$ - we have ruled out such a scenario in our simulations with \approxaii{}  (Fig.~\ref{fig:approxa64_ashes}): we notice that material does not accumulates at $A = 64$ but still peaks at $A = 56$. 
% - - - - - - - - - - - - - - - - - - - - - - - - - - - - - - - - - - - - - - - - - - - - - - - - - - - - - %

%%--%%--%%--%%--%%--%%--%%--%%--%%--%%--%%--%%--%%--%%--%%--%%--%%--%%--%%--%%--%%--%%--%%--%%--%%--%%--%%--%%--%%--%%--%%--%%--%%--%%--%%--%%--%%--%%--%%--%%--%%--%%--%%--%%--%%--%%--%%--%%--%%--%%--%%--%%--%%--%%--%%--%%--%%
%%--%%--%%--%%--%%--%%--%%--%%--%%--%%--%%--%%--%%--%%--%%--%%--%%--%%--%%--%%--%%--%%--%%--%%--%%--%%--%%--%%--%%--%%--%%--%%--%%--%%--%%--%%--%%--%%--%%--%%--%%--%%--%%--%%--%%--%%--%%--%%--%%--%%--%%--%%--%%--%%--%%--%%--%%

%%--%%--%%--%%--%%--%%--%%--%%--%%--%%--%%--%%--%%--%%--%%--%%--%%--%%--%%--%%--%%--%%--%%--%%--%%--%%--%%--%%--%%--%%--%%--%%--%%--%%--%%--%%--%%--%%--%%--%%--%%--%%--%%--%%--%%--%%--%%--%%--%%--%%--%%--%%--%%--%%--%%--%%--%%
%%--%%--%%--%%--%%--%%--%%--%%--%%--%%--%%--%%--%%--%%--%%--%%--%%--%%--%%--%%--%%--%%--%%--%%--%%--%%--%%--%%--%%--%%--%%--%%--%%--%%--%%--%%--%%--%%--%%--%%--%%--%%--%%--%%--%%--%%--%%--%%--%%--%%--%%--%%--%%--%%--%%--%%--%%
\subsection{On the superburst light curve}\label{subsec:discussion_3}
%%--%%--%%--%%--%%--%%--%%--%%--%%--%%--%%--%%--%%--%%--%%--%%--%%--%%--%%--%%--%%--%%--%%--%%--%%--%%--%%--%%--%%--%%--%%--%%--%%--%%--%%--%%--%%--%%--%%--%%--%%--%%--%%--%%--%%--%%--%%--%%--%%--%%--%%--%%--%%--%%--%%--%%--%%

% - - - - - - - - - - - - - - - - - - - - - - - - - - - - - - - - - - - - - - - - - - - - - - - - - - - - - %
Aiming at explaining the 4U 1636$-$53 superburst luminosity curve (see Fig. 15 in KH11), it was shown that at a fixed mass accretion rate comparable to our \middmdot, the base luminosity had an effect over the emergence of a precursor explosion and the decay time but not over the peak luminosity of the superburst. By inspection of our luminosity curves in Fig.~\ref{fig:luminosity_profiles_1}, we notice that modifying the carbon fusion rate at $T\leq 10^{9}$~K has a comparable effect as to increase $L_b$ - or $Q_b$. This suggests that having an enhanced fusion of carbon at low temperatures - THM-simulation - might alleviate the necessity of invoking high amounts of base heating at the moment of fitting the numerical models with observations, as long as the mass accretion rate is well constrained. If, however, the mass accretion rate is not well constrained - as is the case for the 4U 1636$-$53 source - there exists room for considering the HIN-rate to be the actual one for carbon burning, albeit at the cost of invoking a high base luminosity. 
Nevertheless, both our calculations and those of KH11 reproduce the observed superburst properties with base heating values below 0.5 MeV per baryon, suggesting that substantially larger values may not be required by current multi-zone models.
% - - - - - - - - - - - - - - - - - - - - - - - - - - - - - - - - - - - - - - - - - - - - - - - - - - - - - %

% - - - - - - - - - - - - - - - - - - - - - - - - - - - - - - - - - - - - - - - - - - - - - - - - - - - - - %
It has also been suggested that superburst light curves might be used to constrain the presence of low-lying resonances affecting the carbon fusion rate. This argument is based on the fact that a low-temperature resonance in the carbon fusion rate demands less extreme conditions of temperature and pressure for explosions to take place \citep{Cooper_2009,Bravo_2011, Chieffi:2025hpj}.
In this regard, we identify a few caveats limiting the strength of such an interpretation. In first place, usual estimates are based in one-zone model criteria for the explosion condition. However, multi-zone simulations have demonstrated that such approximations may overestimate the ignition column depth and temperature, particularly for the well-studied cases of hydrogen and helium burning \citep[see e.g.][]{PC_ozm_2012, Zamfir_2014, Cyburt_2016, Johnston_2018, Galloway_2021}. Our results, contrasted against the predictions from \citet{Meisel_2024}, suggest a similar caution with respect to the triggering conditions for superbursts.
The second caveat in such an argument can be inferred from the KH11 simulations. In their systematic exploration of parameters, albeit with the CF88 reaction, they showed that a change in the base luminosity at a fixed mass accretion rate has a large impact over the ignition depth. For instance, in simulations with the typical accretion rate $0.3\dot{M}_\mathrm{Edd}$, the column depth at ignition varies between $2\times 10^{11}$ up to $10^{12}$ g~cm$^{-2}$ depending on the adopted value of $Q_b$. We find that introducing or removing a low-temperature resonance in the carbon fusion rate produces changes in the ignition depth that are comparable to those obtained by varying the base luminosity at fixed mass accretion rate (e.g. Subsection~\ref{subsec:qb_vs_rate}).
% - - - - - - - - - - - - - - - - - - - - - - - - - - - - - - - - - - - - - - - - - - - - - - - - - - - - - %

% - - - - - - - - - - - - - - - - - - - - - - - - - - - - - - - - - - - - - - - - - - - - - - - - - - - - - %
In our superburst simulations, we find that photodisintegrations play an important role in the nucleosynthesis around the $A = 56$ isobaric group. In particular, carbon fusion alone is not efficient enough to synthesize these nuclides.
This finding brings support to the proposal of SBC03 that photodisintegrations might contribute to the burst energetics.
This means that the nuclear energy release inferred from light-curve fitting may itself become partially degenerate with the assumed energy source. The nuclear energy, however, is one of the two parameters traditionally used to fit superburst light curves \citep[e.g.][]{Cumming_2004, Cumming_2006, intZand_2017, Huang_2026}, together with the column depth at ignition. 
The quantitative importance of the photodisintegration contribution to the burst energetics remains uncertain because in the present calculations we only adopted a simplified metal composition rather than a detailed distribution of rp-process ashes. Therefore, the possible degeneracies arising as a consequence of the photodisintegration of rp-process ashes deserve further investigation.
% - - - - - - - - - - - - - - - - - - - - - - - - - - - - - - - - - - - - - - - - - - - - - - - - - - - - - %

%%--%%--%%--%%--%%--%%--%%--%%--%%--%%--%%--%%--%%--%%--%%--%%--%%--%%--%%--%%--%%--%%--%%--%%--%%--%%--%%--%%--%%--%%--%%--%%--%%--%%--%%--%%--%%--%%--%%--%%--%%--%%--%%--%%--%%--%%--%%--%%--%%--%%--%%--%%--%%--%%--%%--%%--%%
%%--%%--%%--%%--%%--%%--%%--%%--%%--%%--%%--%%--%%--%%--%%--%%--%%--%%--%%--%%--%%--%%--%%--%%--%%--%%--%%--%%--%%--%%--%%--%%--%%--%%--%%--%%--%%--%%--%%--%%--%%--%%--%%--%%--%%--%%--%%--%%--%%--%%--%%--%%--%%--%%--%%--%%--%%

%%%%%%%%%%%%%%%%%%%%%%%%%%%%%%%%%%%%%%%%%%%%%%%%%%%%%%%%%%%%%%%%%%%%%%%%%%%%%%%%%%%%%%%%%%%%%%%%%%%%%%%%%%%%%%%%%%%%%%%%%%%%%%%%%%%%%%%%%%%%%%%%%%%%%%%%%%%%%%%%%%%%%%%%%%%%%%%%%%%%%%%%%%%%%%%%%%%%%%%%%%%%%%%%%%%%%%%%%%%%%%%%%%%
%%%%%%%%%%%%%%%%%%%%%%%%%%%%%%%%%%%%%%%%%%%%%%%%%%%%%%%%%%%%%%%%%%%%%%%%%%%%%%%%%%%%%%%%%%%%%%%%%%%%%%%%%%%%%%%%%%%%%%%%%%%%%%%%%%%%%%%%%%%%%%%%%%%%%%%%%%%%%%%%%%%%%%%%%%%%%%%%%%%%%%%%%%%%%%%%%%%%%%%%%%%%%%%%%%%%%%%%%%%%%%%%%%%

%%%%%%%%%%%%%%%%%%%%%%%%%%%%%%%%%%%%%%%%%%%%%%%%%%%%%%%%%%%%%%%%%%%%%%%%%%%%%%%%%%%%%%%%%%%%%%%%%%%%%%%%%%%%%%%%%%%%%%%%%%%%%%%%%%%%%%%%%%%%%%%%%%%%%%%%%%%%%%%%%%%%%%%%%%%%%%%%%%%%%%%%%%%%%%%%%%%%%%%%%%%%%%%%%%%%%%%%%%%%%%%%%%%
%%%%%%%%%%%%%%%%%%%%%%%%%%%%%%%%%%%%%%%%%%%%%%%%%%%%%%%%%%%%%%%%%%%%%%%%%%%%%%%%%%%%%%%%%%%%%%%%%%%%%%%%%%%%%%%%%%%%%%%%%%%%%%%%%%%%%%%%%%%%%%%%%%%%%%%%%%%%%%%%%%%%%%%%%%%%%%%%%%%%%%%%%%%%%%%%%%%%%%%%%%%%%%%%%%%%%%%%%%%%%%%%%%%
\section{Conclusions}\label{sec:ending}
%%%%%%%%%%%%%%%%%%%%%%%%%%%%%%%%%%%%%%%%%%%%%%%%%%%%%%%%%%%%%%%%%%%%%%%%%%%%%%%%%%%%%%%%%%%%%%%%%%%%%%%%%%%%%%%%%%%%%%%%%%%%%%%%%%%%%%%%%%%%%%%%%%%%%%%%%%%%%%%%%%%%%%%%%%%%%%%%%%%%%%%%%%%%%%%%%%%%%%%%%%%%%%%%%%%%%%%%%%%%%%%%%%%

% - - - - - - - - - - - - - - - - - - - - - - - - - - - - - - - - - - - - - - - - - - - - - - - - - - - - - %
In this present study, we used the stellar evolution code \mesa{} to analyze the role of the carbon fusion rate on superbursts taking place in the crust of neutron stars. For this purpose, besides the standard carbon fusion rate of \cite{CF88_bib}, we considered three recent determinations - namely THM, HIN and HIN-RES - which significantly differ at temperatures below $10^{9}$~K. In addition, we considered two fiducial mass accretion rates which are consistent with either existing numerical simulations of superbursts \citep[e.g.][]{KH2011} or inferred values from observations \citep{intZand_2017}.

Overall, our results with the standard CF88 rate show a good agreement with the available multi-zone calculations of superbursts from \citet{KH2011}, both in the ignition properties and in the resulting distribution of ashes. Therefore, we are confident that both numerical codes, \mesa{} and \kepler, produce similar results for what concerns superbursts.

% - - - - - - - - - - - - - - - - - - - - - - - - - - - - - - - - - - - - - - - - - - - - - - - - - - - - - %

% - - - - - - - - - - - - - - - - - - - - - - - - - - - - - - - - - - - - - - - - - - - - - - - - - - - - - %
Regardless of the mass accretion rate, we find that a decreased carbon fusion rate at low temperatures, as is the case of the HIN rate, leads to an overall increase in recurrence and decay times as well as $y_{\mathrm{ignition}}$ for superburst, with respect to the reference CF88 simulation. When the carbon fusion rate is boosted at $T\leq 10^{9}$~K, as suggested by the THM rate, we find shorter recurrence and decay times with respect to the reference CF88 simulation, as well as a less-dense environment to trigger the explosion, as reported in Table~\ref{tab:table_1_superbursts_parameters_a56} and in Fig.~\ref{fig:ignition_conditions}. In contrast to these cases, employing the HIN-RES rate - a very similar one to the CF88 rate at $T\geq 8\times 10^{8}$~K - leads to predictions which are almost indistinguishable from those with the CF88 rate. These predictions concern not only the timescales and the ignition column depth of the superburst, but also the distribution of the ashes from carbon burning.
However, we can also affirm that solely modifying the reaction rate at low temperatures cannot mimic the effect of changing the mass accretion rate of the simulation.
In particular, the recurrence time in Table~\ref{tab:table_1_superbursts_parameters_a56} remains of the same order of magnitude independently of the reaction rate at a fixed $\dot{M}$. This implies that a CF88-simulation at \highmdot cannot be simulated by the THM rate at \middmdot, although such a conclusion is limited to the event in which we observe two superbursts in the same source. 

Similarly to KH11 we find that using base heating luminosity values of $\leq 0.5$ MeV per baryon, at different mass accretion rate and regardless of the carbon fusion rate version, is sufficient for triggering superbursts at $y\leq 10^{12}$ g cm$^{2}$. These findings suggest that the required net amount of deep crustal heating is one order of magnitude smaller than what is usually expected, in particular in connection with the shallow heating picture \citep{Meisel_2024}.
However, we also find that variations in the carbon fusion rate produce effects on the ignition depth and light-curve timescales comparable to those obtained by varying the base heating $L_b$. Since the base luminosity cannot be independently constrained, our results suggest that carbon fusion rate uncertainties and envelope heating conditions might not be discernible from solely fitting superburst light curves.
% - - - - - - - - - - - - - - - - - - - - - - - - - - - - - - - - - - - - - - - - - - - - - - - - - - - - - %

% - - - - - - - - - - - - - - - - - - - - - - - - - - - - - - - - - - - - - - - - - - - - - - - - - - - - - %
The superburst ignition conditions obtained here, as well as those in KH11, support the interpretation of superbursts as events taking place in the thermonuclear and not the pycnonuclear regime, the latter corresponding to $\rho\sim 10^{9}$ g cm$^{-3}$ and $T\leq 10^{8}$ K \citep[e.g.][]{Rauscher_2020, Shchechilin_2021, Shchechilin_2022}. 
Notice that adopting different versions of the carbon fusion rate does not change the classification of superbursts as thermonuclear rather than pycnonuclear. However, the resulting differences in the ignition conditions and superbursts properties suggest that a better determination of the low-energy carbon fusion rate could help to reduce one of the sources of uncertainty in theoretical models of superbursts.
% - - - - - - - - - - - - - - - - - - - - - - - - - - - - - - - - - - - - - - - - - - - - - - - - - - - - - %

% - - - - - - - - - - - - - - - - - - - - - - - - - - - - - - - - - - - - - - - - - - - - - - - - - - - - - %
While we have shown that carbon burning at \highmdot might lead to an enhancement in the amount of $^{16}$O, unstable burning of this isotope is unlikely to lead to an explosion since the average mass fraction is far below $1\%$ of the total material. However, it is interesting to notice that a large fraction of material ($\sim 10\%$) goes into \silicos, an isotope which upon electron captures might decay into $^{28}$Ne if pushed at a column depth of $\sim 10^{11}$~g~cm$^{-3}$. Given that this distribution of ashes takes place at \highmdot, it remains to be verified whether such a trend is common to models with $\dot{M}> $ \highmdot; this would represent an argument in favor of neon as the seed element for a hyperburst \citep{Page_2022}, an explosion expected to take place at $\sim 10^{11}$ g cm$^{-3}$ and at such high mass accretion rates. These subsequent chains of reactions over the superburst ashes, however, correspond to the pycnonuclear regime, which deserves further considerations.
% - - - - - - - - - - - - - - - - - - - - - - - - - - - - - - - - - - - - - - - - - - - - - - - - - - - - - %

% - - - - - - - - - - - - - - - - - - - - - - - - - - - - - - - - - - - - - - - - - - - - - - - - - - - - - %
Superbursts are assumed to be triggered by the burning of the ashes from the rp-process. These ashes have been shown to contain nuclides in the range $A = 56$ up to $A\sim 100$ \citep{Meisel_2018}.
The final products of superbursts remain to be calculated within a multi-zone code when the full distribution of ashes, instead of the fiducial $80\%$ of \iroh, is taken into account. Furthermore, whether superbursts can be triggered with a different mass fraction of \carbos{} - e.g. below the canonical $20\%$ value should be further investigated.
% - - - - - - - - - - - - - - - - - - - - - - - - - - - - - - - - - - - - - - - - - - - - - - - - - - - - - %

%%%%%%%%%%%%%%%%%%%%%%%%%%%%%%%%%%%%%%%%%%%%%%%%%%%%%%%%%%%%%%%%%%%%%%%%%%%%%%%%%%%%%%%%%%%%%%%%%%%%%%%%%%%%%%%%%%%%%%%%%%%%%%%%%%%%%%%%%%%%%%%%%%%%%%%%%%%%%%%%%%%%%%%%%%%%%%%%%%%%%%%%%%%%%%%%%%%%%%%%%%%%%%%%%%%%%%%%%%%%%%%%%%%
%%%%%%%%%%%%%%%%%%%%%%%%%%%%%%%%%%%%%%%%%%%%%%%%%%%%%%%%%%%%%%%%%%%%%%%%%%%%%%%%%%%%%%%%%%%%%%%%%%%%%%%%%%%%%%%%%%%%%%%%%%%%%%%%%%%%%%%%%%%%%%%%%%%%%%%%%%%%%%%%%%%%%%%%%%%%%%%%%%%%%%%%%%%%%%%%%%%%%%%%%%%%%%%%%%%%%%%%%%%%%%%%%%%

%%%%%%%%%%%%%%%%%%%%%%%%%%%%%%%%%%%%%%%%%%%%%%%%%%%%%%%%%%%%%%%%%%%%%%%%%%%%%%%%%%%%%%%%%%%%%%%%%%%%%%%%%%%%%%%%%%%%%%%%%%%%%%%%%%%%%%%%%%%%%%%%%%%%%%%%%%%%%%%%%%%%%%%%%%%%%%%%%%%%%%%%%%%%%%%%%%%%%%%%%%%%%%%%%%%%%%%%%%%%%%%%%%%
%%%%%%%%%%%%%%%%%%%%%%%%%%%%%%%%%%%%%%%%%%%%%%%%%%%%%%%%%%%%%%%%%%%%%%%%%%%%%%%%%%%%%%%%%%%%%%%%%%%%%%%%%%%%%%%%%%%%%%%%%%%%%%%%%%%%%%%%%%%%%%%%%%%%%%%%%%%%%%%%%%%%%%%%%%%%%%%%%%%%%%%%%%%%%%%%%%%%%%%%%%%%%%%%%%%%%%%%%%%%%%%%%%%
% ACKNOWLEDGEMENTS
%%%%%%%%%%%%%%%%%%%%%%%%%%%%%%%%%%%%%%%%%%%%%%%%%%%%%%%%%%%%%%%%%%%%%%%%%%%%%%%%%%%%%%%%%%%%%%%%%%%%%%%%%%%%%%%%%%%%%%%%%%%%%%%%%%%%%%%%%%%%%%%%%%%%%%%%%%%%%%%%%%%%%%%%%%%%%%%%%%%%%%%%%%%%%%%%%%%%%%%%%%%%%%%%%%%%%%%%%%%%%%%%%%%
\begin{acknowledgements}
M.N.-C. acknowledges support by the Fonds de la Recherche Scientifique-FNRS under Grant No IISN 4.4502.19. 
The authors are members of BLU-ULB, the interfaculty research group focusing on space research at ULB - Universit\'e Libre de Bruxelles. 

The inlists for the \mesa{} simulations are available under an open-source Creative Commons Attribution license on Zenodo, at DOI \url{https://doi.org/10.5281/zenodo.21341165}\, \citep{nava_callejas_2026_21341165}.

\end{acknowledgements}
%%%%%%%%%%%%%%%%%%%%%%%%%%%%%%%%%%%%%%%%%%%%%%%%%%%%%%%%%%%%%%%%%%%%%%%%%%%%%%%%%%%%%%%%%%%%%%%%%%%%%%%%%%%%%%%%%%%%%%%%%%%%%%%%%%%%%%%%%%%%%%%%%%%%%%%%%%%%%%%%%%%%%%%%%%%%%%%%%%%%%%%%%%%%%%%%%%%%%%%%%%%%%%%%%%%%%%%%%%%%%%%%%%%
%%%%%%%%%%%%%%%%%%%%%%%%%%%%%%%%%%%%%%%%%%%%%%%%%%%%%%%%%%%%%%%%%%%%%%%%%%%%%%%%%%%%%%%%%%%%%%%%%%%%%%%%%%%%%%%%%%%%%%%%%%%%%%%%%%%%%%%%%%%%%%%%%%%%%%%%%%%%%%%%%%%%%%%%%%%%%%%%%%%%%%%%%%%%%%%%%%%%%%%%%%%%%%%%%%%%%%%%%%%%%%%%%%%

%%%%%%%%%%%%%%%%%%%%%%%%%%%%%%%%%%%%%%%%%%%%%%%%%%%%%%%%%%%%%%%%%%%%%%%%%%%%%%%%%%%%%%%%%%%%%%%%%%%%%%%%%%%%%%%%%%%%%%%%%%%%%%%%%%%%%%%%%%%%%%%%%%%%%%%%%%%%%%%%%%%%%%%%%%%%%%%%%%%%%%%%%%%%%%%%%%%%%%%%%%%%%%%%%%%%%%%%%%%%%%%%%%%
%%%%%%%%%%%%%%%%%%%%%%%%%%%%%%%%%%%%%%%%%%%%%%%%%%%%%%%%%%%%%%%%%%%%%%%%%%%%%%%%%%%%%%%%%%%%%%%%%%%%%%%%%%%%%%%%%%%%%%%%%%%%%%%%%%%%%%%%%%%%%%%%%%%%%%%%%%%%%%%%%%%%%%%%%%%%%%%%%%%%%%%%%%%%%%%%%%%%%%%%%%%%%%%%%%%%%%%%%%%%%%%%%%%
% BIBLIOGRAPHY
%%%%%%%%%%%%%%%%%%%%%%%%%%%%%%%%%%%%%%%%%%%%%%%%%%%%%%%%%%%%%%%%%%%%%%%%%%%%%%%%%%%%%%%%%%%%%%%%%%%%%%%%%%%%%%%%%%%%%%%%%%%%%%%%%%%%%%%%%%%%%%%%%%%%%%%%%%%%%%%%%%%%%%%%%%%%%%%%%%%%%%%%%%%%%%%%%%%%%%%%%%%%%%%%%%%%%%%%%%%%%%%%%%%
\bibliographystyle{aa}
\bibliography{aa61527-26_bib}
%%%%%%%%%%%%%%%%%%%%%%%%%%%%%%%%%%%%%%%%%%%%%%%%%%%%%%%%%%%%%%%%%%%%%%%%%%%%%%%%%%%%%%%%%%%%%%%%%%%%%%%%%%%%%%%%%%%%%%%%%%%%%%%%%%%%%%%%%%%%%%%%%%%%%%%%%%%%%%%%%%%%%%%%%%%%%%%%%%%%%%%%%%%%%%%%%%%%%%%%%%%%%%%%%%%%%%%%%%%%%%%%%%%
%%%%%%%%%%%%%%%%%%%%%%%%%%%%%%%%%%%%%%%%%%%%%%%%%%%%%%%%%%%%%%%%%%%%%%%%%%%%%%%%%%%%%%%%%%%%%%%%%%%%%%%%%%%%%%%%%%%%%%%%%%%%%%%%%%%%%%%%%%%%%%%%%%%%%%%%%%%%%%%%%%%%%%%%%%%%%%%%%%%%%%%%%%%%%%%%%%%%%%%%%%%%%%%%%%%%%%%%%%%%%%%%%%%

%%%%%%%%%%%%%%%%%%%%%%%%%%%%%%%%%%%%%%%%%%%%%%%%%%%%%%%%%%%%%%%%%%%%%%%%%%%%%%%%%%%%%%%%%%%%%%%%%%%%%%%%%%%%%%%%%%%%%%%%%%%%%%%%%%%%%%%%%%%%%%%%%%%%%%%%%%%%%%%%%%%%%%%%%%%%%%%%%%%%%%%%%%%%%%%%%%%%%%%%%%%%%%%%%%%%%%%%%%%%%%%%%%%
%%%%%%%%%%%%%%%%%%%%%%%%%%%%%%%%%%%%%%%%%%%%%%%%%%%%%%%%%%%%%%%%%%%%%%%%%%%%%%%%%%%%%%%%%%%%%%%%%%%%%%%%%%%%%%%%%%%%%%%%%%%%%%%%%%%%%%%%%%%%%%%%%%%%%%%%%%%%%%%%%%%%%%%%%%%%%%%%%%%%%%%%%%%%%%%%%%%%%%%%%%%%%%%%%%%%%%%%%%%%%%%%%%%
\begin{appendix}
%%%%%%%%%%%%%%%%%%%%%%%%%%%%%%%%%%%%%%%%%%%%%%%%%%%%%%%%%%%%%%%%%%%%%%%%%%%%%%%%%%%%%%%%%%%%%%%%%%%%%%%%%%%%%%%%%%%%%%%%%%%%%%%%%%%%%%%%%%%%%%%%%%%%%%%%%%%%%%%%%%%%%%%%%%%%%%%%%%%%%%%%%%%%%%%%%%%%%%%%%%%%%%%%%%%%%%%%%%%%%%%%%%%

%%--%%--%%--%%--%%--%%--%%--%%--%%--%%--%%--%%--%%--%%--%%--%%--%%--%%--%%--%%--%%--%%--%%--%%--%%--%%--%%--%%--%%--%%--%%--%%--%%--%%--%%--%%--%%--%%--%%--%%--%%--%%--%%--%%--%%--%%--%%--%%--%%--%%--%%--%%--%%--%%--%%--%%--%%
%%--%%--%%--%%--%%--%%--%%--%%--%%--%%--%%--%%--%%--%%--%%--%%--%%--%%--%%--%%--%%--%%--%%--%%--%%--%%--%%--%%--%%--%%--%%--%%--%%--%%--%%--%%--%%--%%--%%--%%--%%--%%--%%--%%--%%--%%--%%--%%--%%--%%--%%--%%--%%--%%--%%--%%--%%
\section{Calculation of the mass fraction averages}\label{appendix:mass_fraction_computation}
%%--%%--%%--%%--%%--%%--%%--%%--%%--%%--%%--%%--%%--%%--%%--%%--%%--%%--%%--%%--%%--%%--%%--%%--%%--%%--%%--%%--%%--%%--%%--%%--%%--%%--%%--%%--%%--%%--%%--%%--%%--%%--%%--%%--%%--%%--%%--%%--%%--%%--%%--%%--%%--%%--%%--%%--%%

%----------------------------------------------------------------------------------------------------------------------------------%
\begin{figure}[ht!]
    \centering
    \includegraphics[width=0.40\textwidth]{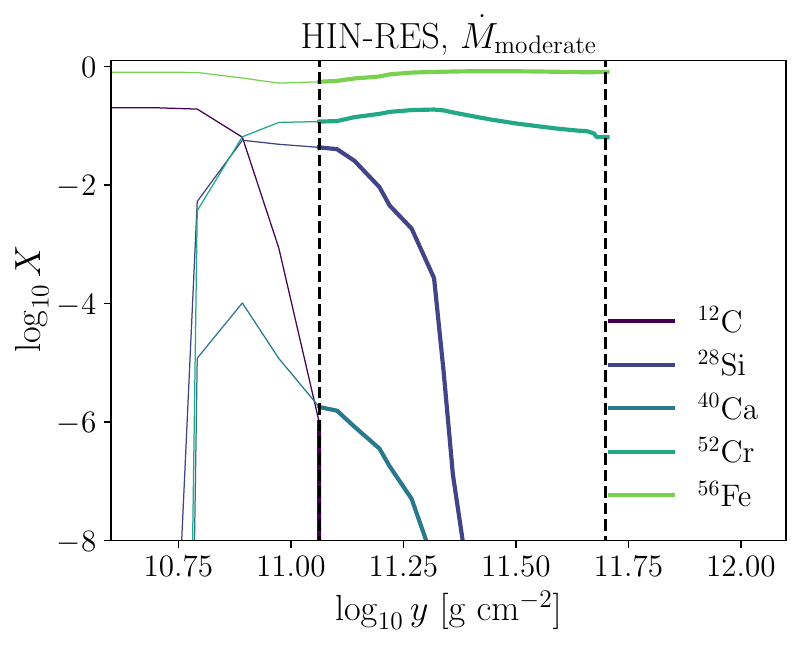}
    \caption{Mass fractions of a selected set of nuclides as a function of column depth for the HIN-RES simulation at \middmdot, 20 hrs after the luminosity peak of the superburst. The dashed vertical lines illustrate the column depths between which the mass fraction average is computed according to Eq.~\ref{eq:mass_average_1}.}
    \label{fig:mass_fractions_for_average_example}
\end{figure}
%----------------------------------------------------------------------------------------------------------------------------------%

While in multi-zone simulations of a burst or superburst the ignition takes place at a rather well-defined column depth $y_{\mathrm{ignition}}$, the fuel burnt during the explosion is spread along a region $\Delta y$. As a consequence, ``the chemical compositon of the ashes from a burst/superburst'' properly refers to a region of the envelope in the column depth coordinate instead of a fixed location. Moreover, both the ignition depth and the extension of the ashes are sensitive to the mass accretion rate. To get a visual example of this situation, in Fig.~\ref{fig:mass_fractions_for_average_example} we show the chemical distribution of a few isotopes as a function of column depth for the HIN-RES simulation at \middmdot.

Consequently, in order to perform a fair comparison of mass fractions or abundances of the ashes from a burst it is instructive to compute and analyze, for each nuclide in the network, their individual average mass fraction over $\Delta y$, defined as
\begin{equation}
\langle X_{i}\rangle := \frac{\int_{\Delta y}dy\, X_{i}(y)}{\int_{\Delta y}dy}.
\label{eq:mass_average_1}
\end{equation}
where $X(y)$ is the mass fraction of isotope $i$ as a function of the column depth. Once the individual mass fractions are computed we can get the total average mass fraction as a function of $Z$ or $A$ of the ashes by adding up all the averages of isotopes with same $Z$ or $A$. 
For Figs.~\ref{fig:distribution_ashess_1} and~\ref{fig:distribution_ashess_1b} we take for $\Delta y$ the region ranging between the column depth $y_c$ at which the mass fraction of \carbos{} is $10^{-6}$ and the maximum column depth reached by accreted material at an instant $t$, i.e. $y_{\mathrm{max}} = \dot{m}t$ with $\dot{m} = \dot{M}/(4\pi R^{2})$. Here we assume $R$ to be the radius of the underlying neutron star, $\dot{M}$ the mass accretion rate and $t$ the time at which the average is computed. For the averages displayed in Fig.~\ref{fig:mass_fractions_1} we opt for taking as the lower limit of integration the column depth $y = 10^{9}$~g~cm$^{-2}$, so that only the fuel layers close to the ignition point are included in the nucleosynthesis discussion, and not those close to the surface where the composition is mainly dictated by the accreted material, e.g. $20\%$ \carbos{} and $80\%$ \iroh.

%%--%%--%%--%%--%%--%%--%%--%%--%%--%%--%%--%%--%%--%%--%%--%%--%%--%%--%%--%%--%%--%%--%%--%%--%%--%%--%%--%%--%%--%%--%%--%%--%%--%%--%%--%%--%%--%%--%%--%%--%%--%%--%%--%%--%%--%%--%%--%%--%%--%%--%%--%%--%%--%%--%%--%%--%%
%%--%%--%%--%%--%%--%%--%%--%%--%%--%%--%%--%%--%%--%%--%%--%%--%%--%%--%%--%%--%%--%%--%%--%%--%%--%%--%%--%%--%%--%%--%%--%%--%%--%%--%%--%%--%%--%%--%%--%%--%%--%%--%%--%%--%%--%%--%%--%%--%%--%%--%%--%%--%%--%%--%%--%%--%%

%%--%%--%%--%%--%%--%%--%%--%%--%%--%%--%%--%%--%%--%%--%%--%%--%%--%%--%%--%%--%%--%%--%%--%%--%%--%%--%%--%%--%%--%%--%%--%%--%%--%%--%%--%%--%%--%%--%%--%%--%%--%%--%%--%%--%%--%%--%%--%%--%%--%%--%%--%%--%%--%%--%%--%%--%%
%%--%%--%%--%%--%%--%%--%%--%%--%%--%%--%%--%%--%%--%%--%%--%%--%%--%%--%%--%%--%%--%%--%%--%%--%%--%%--%%--%%--%%--%%--%%--%%--%%--%%--%%--%%--%%--%%--%%--%%--%%--%%--%%--%%--%%--%%--%%--%%--%%--%%--%%--%%--%%--%%--%%--%%--%%
\section{On the impact of the \carbos{} + \carbos{} $n-$channel for the superburst simulations}\label{appendix:n_channel_supeburst}
%%--%%--%%--%%--%%--%%--%%--%%--%%--%%--%%--%%--%%--%%--%%--%%--%%--%%--%%--%%--%%--%%--%%--%%--%%--%%--%%--%%--%%--%%--%%--%%--%%--%%--%%--%%--%%--%%--%%--%%--%%--%%--%%--%%--%%--%%--%%--%%--%%--%%--%%--%%--%%--%%--%%--%%--%%

%----------------------------------------------------------------------------------------------------------------------------------%
\begin{figure}[ht!]
    \centering
    \includegraphics[width=0.4\textwidth]{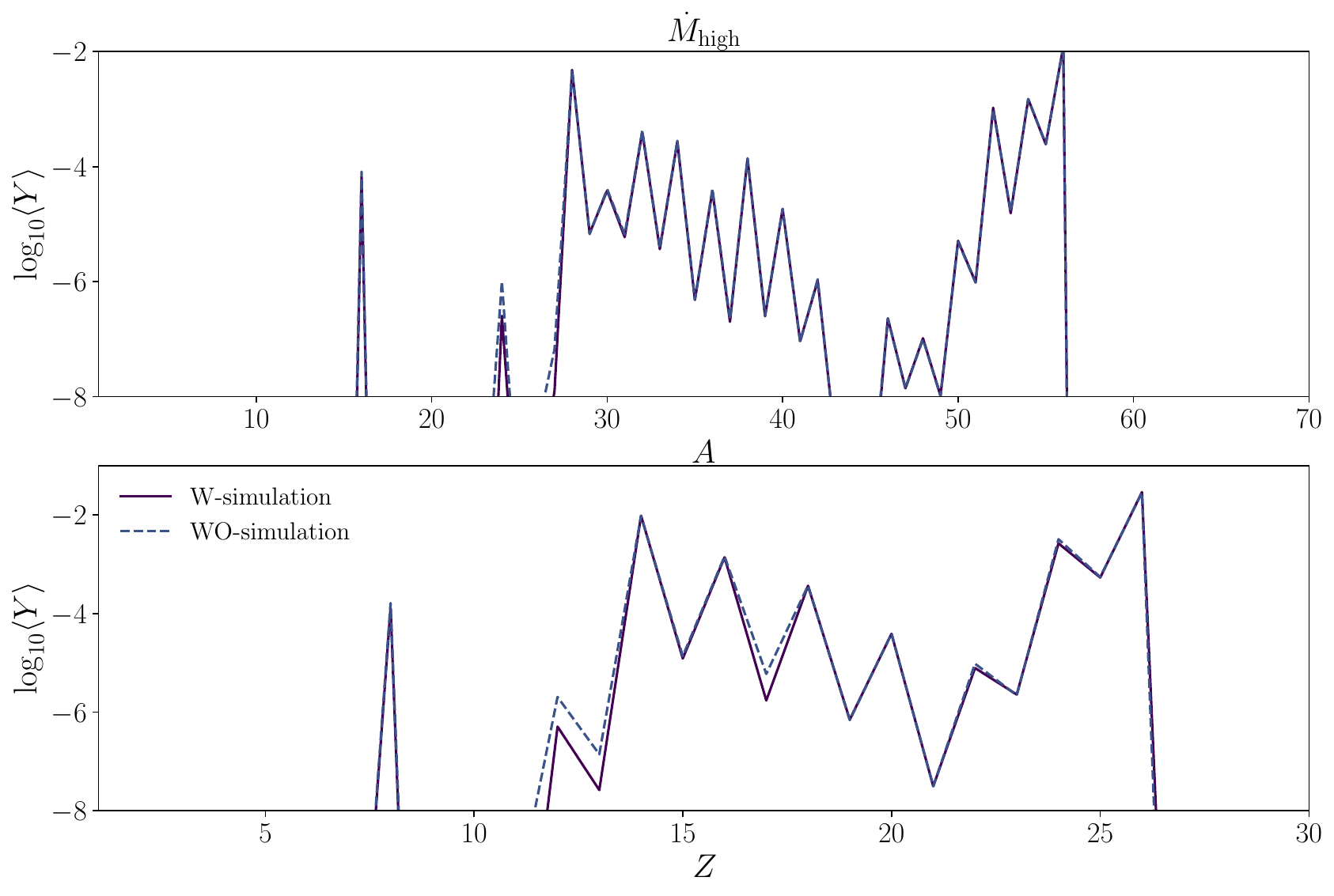}
    \caption{Same caption as in Fig.~\ref{fig:distribution_ashess_1}, now for two models at \highmdot with, W, and without, WO, the $n-$channel of the carbon fusion rate.}
    \label{fig:distribution_ashess_appendixb}
\end{figure}
%----------------------------------------------------------------------------------------------------------------------------------%

In Sect.~\ref{sec:results} we saw that the maximum temperature of the envelope during a superburst exceeds $10^{9}$~K. According to CF88, at these temperatures, the $n$-channel of the carbon fusion rate takes up to $\sim 10\%$ from the total rate, and thus its effects might be of importance although, at first glance, the setup of the superburst is not a site enriched in neutrons. However, since in the present work we have modified only the $p-$ and $\alpha-$channels, it is important to test the impact of the $n-$channel. In order to do so, we ran a \mesa\ simulation at \highmdot, with the \approxai\ network and with the CF88 versions for the $p-$ and $\alpha-$ channels but removing both the $n-$channel and its inverse reaction, hereafter WO-simulation, and in this Appendix we compare it against the simulation including the $n-$ channel and its inverse rate - hereafter W-simulation. In both cases, three superbursts are followed.

With respect to the maximum temperature of the envelope and its evolution we find no discrepancy between W- and WO-simulations. This is also the case for the luminosity decay time, which in both simulations is of $ 0.40$ hr. Regarding the luminosity peaks of superbursts, the recurrence time between events and the column depth at ignition we observe discrepancies inferior to $3\%$. 

The major discrepancy we find is in the chemical composition of the ashes. In Fig.~\ref{fig:distribution_ashess_appendixb} we compare the distribution of ashes as a function of $A$ and $Z$ for W- and WO-simulations: we notice removing the $n-$channel leads to a slight overproduction of $^{23}$Mg. However, the rest of distribution is fairly similar.

From these comparisons we conclude the presence of the $n-$channel and the absence of a modification for this rate and its inverse does not have a major effect in the findings of the present work.

%%--%%--%%--%%--%%--%%--%%--%%--%%--%%--%%--%%--%%--%%--%%--%%--%%--%%--%%--%%--%%--%%--%%--%%--%%--%%--%%--%%--%%--%%--%%--%%--%%--%%--%%--%%--%%--%%--%%--%%--%%--%%--%%--%%--%%--%%--%%--%%--%%--%%--%%--%%--%%--%%--%%--%%--%%
%%--%%--%%--%%--%%--%%--%%--%%--%%--%%--%%--%%--%%--%%--%%--%%--%%--%%--%%--%%--%%--%%--%%--%%--%%--%%--%%--%%--%%--%%--%%--%%--%%--%%--%%--%%--%%--%%--%%--%%--%%--%%--%%--%%--%%--%%--%%--%%--%%--%%--%%--%%--%%--%%--%%--%%--%%

\end{appendix}

%%%%%%%%%%%%%%%%%%%%%%%%%%%%%%%%%%%%%%%%%%%%%%%%%%%%%%%%%%%%%%%%%%%%%%%%%%%%%%%%%%%%%%%%%%%%%%%%%%%%%%%%%%%%%%%%%%%%%%%%%%%%%%%%%%%%%%%%%%%%%%%%%%%%%%%%%%%%%%%%%%%%%%%%%%%%%%%%%%%%%%%%%%%%%%%%%%%%%%%%%%%%%%%%%%%%%%%%%%%%%%%%%%%
%%%%%%%%%%%%%%%%%%%%%%%%%%%%%%%%%%%%%%%%%%%%%%%%%%%%%%%%%%%%%%%%%%%%%%%%%%%%%%%%%%%%%%%%%%%%%%%%%%%%%%%%%%%%%%%%%%%%%%%%%%%%%%%%%%%%%%%%%%%%%%%%%%%%%%%%%%%%%%%%%%%%%%%%%%%%%%%%%%%%%%%%%%%%%%%%%%%%%%%%%%%%%%%%%%%%%%%%%%%%%%%%%%%

%%%%%%%%%%%%%%%%%%%%%%%%%%%%%%%%%%%%%%%%%%%%%%%%%%%%%%%%%%%%%%%%%%%%%%%%%%%%%%%%%%%%%%%%%%%%%%%%%%%%%%%%%%%%%%%%%%%%%%%%%%%%%%%%%%%%%%%%%%%%%%%%%%%%%%%%%%%%%%%%%%%%%%%%%%%%%%%%%%%%%%%%%%%%%%%%%%%%%%%%%%%%%%%%%%%%%%%%%%%%%%%%%%%%%%%%%%%%%%%%%%%%%%%%%%%%%%%%%%%%%%%%%%%%%%%%%%%%%%%%%%%%%%%%%%%%%%%%%%%%%%%%%%%%%%%%%%%%%%%%%%%%%%%%%%%%%%%%%%%%%%%%%%%%%%%%%%%%%%%%%%%%%%%%%%%%%%%%%%%%%%%%%%%%%%%%%%%%%%%%%%%%%%%%%%%%%%%%%%%%%%%%%%%%%%%%%%
%%%%%%%%%%%%%%%%%%%%%%%%%%%%%%%%%%%%%%%%%%%%%%%%%%%%%%%%%%%%%%%%%%%%%%%%%%%%%%%%%%%%%%%%%%%%%%%%%%%%%%%%%%%%%%%%%%%%%%%%%%%%%%%%%%%%%%%%%%%%%%%%%%%%%%%%%%%%%%%%%%%%%%%%%%%%%%%%%%%%%%%%%%%%%%%%%%%%%%%%%%%%%%%%%%%%%%%%%%%%%%%%%%%%%%%%%%%%%%%%%%%%%%%%%%%%%%%%%%%%%%%%%%%%%%%%%%%%%%%%%%%%%%%%%%%%%%%%%%%%%%%%%%%%%%%%%%%%%%%%%%%%%%%%%%%%%%%%%%%%%%%%%%%%%%%%%%%%%%%%%%%%%%%%%%%%%%%%%%%%%%%%%%%%%%%%%%%%%%%%%%%%%%%%%%%%%%%%%%%%%%%%%%%%%%%%%%
\end{document}